\documentclass[aps, prx, reprint, floatfix, superscriptaddress]{revtex4-2} 

\usepackage{graphicx}

\usepackage{epstopdf}
\usepackage{bm, amsmath, amssymb}
\usepackage{float}
\usepackage{placeins}
\usepackage{xcolor}
\usepackage{siunitx}
\usepackage[T1]{fontenc}
\usepackage{array,booktabs}
\usepackage{comment}
\usepackage{multirow}
\usepackage{enumitem}

\usepackage{hyperref}
\hypersetup{
    colorlinks=true,
    linkcolor=magenta,
    urlcolor=magenta,
    filecolor=magenta,   
    citecolor=blue
}

\newcommand{\figwidth}{3.375in} 
\newcommand{\figwidthDouble}{7in}

\usepackage{appendix}
\usepackage[resetlabels]{multibib}
\usepackage{titlesec}

\newcites{SM}{SM_refs}

\newcommand{\suppmat}{\appendix}

\makeatletter
\newcommand\suppmat@section[1]{%
  \refstepcounter{section}%
  \orig@section*{Supplementary material \@Alph\c@section: #1}%
  \addcontentsline{toc}{section}{Supplementary material \@Alph\c@section: #1}%
}
\let\orig@section\section
\g@addto@macro\suppmat{\let\section\suppmat@section}
\makeatother

\makeatletter
\newcommand\appendix@section[1]{%
  \refstepcounter{section}%
  \orig@section*{Appendix \@Alph\c@section: #1}%
  \addcontentsline{toc}{section}{Appendix \@Alph\c@section: #1}%
}
\let\orig@section\section
\g@addto@macro\appendix{\let\section\appendix@section}
\makeatother

\begin{document}

\title{Model-Free Quantum Control with Reinforcement Learning}

\author{V. V. Sivak}
\email{vladimir.sivak@yale.edu}
\affiliation{Department of Applied Physics, Yale University, New Haven, CT 06520, USA}

\author{A. Eickbusch}
\affiliation{Department of Applied Physics, Yale University, New Haven, CT 06520, USA}

\author{H. Liu}
\affiliation{Department of Applied Physics, Yale University, New Haven, CT 06520, USA}

\author{B. Royer}
\affiliation{Department of Physics, Yale University, New Haven, CT 06520, USA}

\author{I. Tsioutsios}
\affiliation{Department of Applied Physics, Yale University, New Haven, CT 06520, USA}

\author{M. H. Devoret}
\affiliation{Department of Applied Physics, Yale University, New Haven, CT 06520, USA}

\begin{abstract}
Model bias is an inherent limitation of the current dominant approach to optimal quantum control, which relies on a system simulation for optimization of control policies. To overcome this limitation, we propose a circuit-based approach for training a reinforcement learning agent on quantum control tasks in a  model-free way. Given a continuously parameterized control circuit, the agent learns its parameters through trial-and-error interaction with the quantum system, using measurement outcomes as the only source of information about the quantum state. Focusing on control of a harmonic oscillator coupled to an ancilla qubit, we show how to reward the learning agent using measurements of experimentally available observables. We train the agent to prepare various non-classical states using both unitary control and control with adaptive measurement-based quantum feedback, and to execute logical gates on encoded qubits.
This approach significantly outperforms widely used model-free methods in terms of sample efficiency.
Our numerical work is of immediate relevance to superconducting circuits and trapped ions platforms where such training can be implemented in experiment, allowing complete elimination of model bias and the adaptation of quantum control policies to the specific system in which they are deployed.
\end{abstract}

\maketitle

\section{Introduction}
Quantum control theory addresses a problem of optimally implementing a desired quantum operation using external controls. The design of experimental control policies is currently dominated by {\sl simulation-based} optimal control theory methods with favorable convergence properties thanks to the availability of analytic gradients \cite{Khaneja2005, Caneva2011, DeFouquieres2011} or automatic differentiation \cite{Leung2017, Abdelhafez2019}. However, it is important to acknowledge that simulation-based methods can only be as good as the underlying models used in the simulation. Empirically, model bias leads to a significant degradation of performance of the quantum control policies, when optimized in simulation and then tested in experiment \cite{Kelly2014, Chen2016, Rol2017, Werninghaus2021}. A practical model-free alternative to simulation-based methods in quantum control is thus desirable.

The idea of using model-free optimization in quantum control can be traced back to the pioneering proposal in 1992 of laser pulse shaping for molecular control with a genetic algorithm \cite{Judson1992}. Only in recent years has the controllability of quantum systems and the duty cycle of  optimization feedback loops reached sufficient levels to allow for the experimental implementation of such ideas. The few existing demonstrations are based on model-free optimization algorithms such as Nelder-Mead simplex search \cite{Kelly2014, Chen2016, Rol2017}, evolutionary strategies \cite{Werninghaus2021} and particle swarm optimization \cite{Lumino2018}.

At the same time, deep reinforcement learning (RL) \cite{Sutton2017, Francois-Lavet2018} emerged as not only a powerful optimization technique but also a tool for discovering adaptive decision-making policies. In this framework, learning proceeds by trial-and-error, without access to the model generating the dynamics and its gradients.
Being intrinsically free of model bias, it is an attractive alternative to traditional simulation-based approaches in quantum control. In a variety of domains, deep reinforcement learning has recently produced spectacular results, such as beating world champions in board games \cite{Silver2016, Silver2018}, reaching human-level performance in sophisticated computer games \cite{Mnih2015, Vinyals2019}, and controlling robotic locomotion \cite{Levine2015, Haarnoja2018b}. 

Applying model-free RL to quantum control implies direct interaction of the learning agent with the controlled quantum system, which presents a number of unique challenges not typically encountered in classical environments. Quantum systems have large continuous state spaces that are only partially observable to the agent through measurements. For example, a pure qubit state can be described as a point on a Bloch sphere, but a projective measurement of a qubit observable yields a random {\it binary}$\,$ outcome. Qubits are often used as ancillary systems to control harmonic oscillators, in which case the underlying state space is formally infinite-dimensional. Learning quantum control of such systems is akin to learning to drive a car with a single sensor that provides binary-valued feedback.
The question arises: can classical model-free RL agents efficiently handle such ``quantum-observable'' environments?

The previous applications of RL to quantum control 
\cite{Chen2014, Bukov2018, Bukov2018a, Zhang2019, Zhang2019, Porotti2019, An2020, August2018, Haug2020, Kuo2021, Wang2019a, Borah2021, Dalgaard2019, Niu2019, An2019, Fosel2018, Andreasson2019, Nautrup2019, DomingoColomer2020, Xu2019, Schuff2020}, which we survey in Section~\ref{sec:related work},
relied on a number of simplifying assumptions rendering the quantum control problem more tractable for the agent, but severely limiting their experimental feasibility. 
These approaches provide the agent with the knowledge of a quantum state, or rely on fidelity as a measure of optimization progress. Such requirements are at odds with the fundamental properties of quantum environments, stochasticity and minimalistic observability. 
Trying to meet these requirements in realistic experiments leads to large sample size, e.g. $10^7$ measurements to learn a single-qubit gate with only 16 parameters, as recently demonstrated in Ref.~\cite{Baum2021} using a quantum-state-aware agent that relied on tomography to obtain the quantum state. 
Other model-free approaches that view quantum control as a standard cost function optimization problem \cite{Judson1992, Kelly2014, Chen2016, Rol2017, Werninghaus2021} are subject to similar limitations. Scaling such methods beyond one or two-qubit applications is prohibitively expensive from a practical point of view. 

In this paper, we develop a framework for model-free learning of quantum control policies which is explicitly tailored to the stochasticity and minimalistic quantum observability. It does not rely on restrictive assumptions, such as a model of the system's dynamics, knowledge of a quantum state, or access to fidelity. 
By framing quantum control as a Quantum-Observable Markov decision process (QOMDP) \cite{Barry2014}, we consider each stochastic experimental realization as an episode of interaction of the learning agent with a controlled quantum system, after which the agent receives a binary-valued reward through a projective measurement. Instead of utilizing averaging, every such episode is performed with a {\it different} control policy, which is being continually updated by a small amount within a trust region with the help of the reward signal. This novel policy space exploration strategy leads to excellent  sample efficiency on challenging high-dimensional tasks, significantly outperforming widely used model-free methods.

To illustrate our approach with specific examples, we focus on the quantum control of a harmonic oscillator. Harmonic oscillators are ubiquitous physical systems, realized, for instance, as the motional degrees of freedom of trapped ions \cite{Leibfried2003, Bruzewicz2019} or electromagnetic modes in superconducting circuits \cite{Krantz2019a, Blais2020}. They are primitives for bosonic quantum error correction \cite{Ofek2016a, Campagne-Ibarcq2019, Hu2019a} and quantum sensing \cite{Wang2019b}. Universal quantum control of an oscillator is typically realized by coupling it to an ancillary nonlinear system, such as a qubit, with state-of-the-art fidelities in the $0.9-0.99$ range in circuit quantum electrodynamics (QED) \cite{Heeres2017, Eickbusch2021, Kudra2021} and trapped ions \cite{Fluhmann2019a}. In such a quantum environment, ancilla measurements with binary outcomes are the agent's only source of information about the quantum state in the vast unobservable Hilbert space and the only source of rewards guiding the learning algorithm.

For an oscillator-qubit system, we demonstrate learning of both unitary control and control with adaptive measurement-based quantum feedback. These types of control are special instances of a modular circuit-based framework, in which the quantum operation executed on a system is represented as a sequence of continuously parameterized {\it control circuits}, whose parameters are learned in-situ with the help of a {\it reward circuit}.
We show how to construct task-specific reward circuits that implement an experimentally feasible dichotomic positive operator-valued measure (POVM) on the oscillator, and how to use its outcomes as reward bits in the classical training loop. 
We train the agent to prepare various non-classical oscillator states, such as Fock states, Gottesman-Kitaev-Preskill (GKP) states \cite{Gottesman2001}, Schr\"{o}dinger cat states, and binomial code states \cite{Michael2016}, and to execute gates on logical qubits encoded in an oscillator. 

Although our demonstration is based on a simulated environment producing mock measurement outcomes, the RL agent that we developed (code available at \cite{Sivak2021}) is compatible with real-world experiments.

\section{Related work\label{sec:related work}}
In recent years, multiple theoretical proposals have emerged around applying reinforcement learning to quantum control problems such as quantum state preparation \cite{Chen2014, Bukov2018, Bukov2018a, Zhang2019, Zhang2019, Porotti2019, An2020, August2018, Haug2020, Kuo2021} and feedback stabilization \cite{Wang2019a, Borah2021}, the construction of quantum gates \cite{Dalgaard2019, Niu2019, An2019},  design of quantum error correction protocols \cite{Fosel2018, Andreasson2019, Nautrup2019, DomingoColomer2020}, and control-enhanced quantum sensing \cite{Xu2019, Schuff2020}. 
These proposals formulate the control problem in a way that avoids directly facing quantum observability and makes it more tractable for the RL agent.
In simulated environments, this is possible, for example, by providing the agent with full knowledge of the system's quantum state, which supplies enough information for decision making \cite{Chen2014, Zhang2019, Porotti2019, An2020, Haug2020, Schuff2020, Wang2019a, Fosel2018, Xu2019}. Moreover, in the simulation the distance to the target state or operation is known at every step of the quantum trajectory, and it can be used to construct a steady reward signal to guide the learning algorithm \cite{Zhang2019, Porotti2019, An2020, Xu2019}, thereby alleviating the well-known delayed reward assignment problem \cite{Sutton2017, Francois-Lavet2018}. 
Taking RL a step closer towards quantum observability, some works only provide the agent with access to fidelities and expectation values of physical observables in different parts of the training pipeline \cite{Bukov2018, August2018, Garcia-Saez2019, Wauters2020, Kuo2021}, which would still require a prohibitive amount of averaging in an experiment, a problem exacerbated by the iterative nature of the training process.
Under these various simplifications, there are positive indications \cite{Zhang2019, Dalgaard2019} that RL is able to match the performance of traditional gradient-based methods, albeit in situations where the agent or the learning algorithm has access to expensive or unrealistic resources.
Therefore, such RL proposals are not compatible with efficient training in experiment, which is required in order to eliminate model bias from quantum control. 
To address this challenge, it is necessary to develop agents that learn directly from stochastic measurement outcomes or from low-sample estimators of physical observables. Initial steps towards this goal were studied in \cite{Bukov2018a, Bilkis2020, Borah2021}.

\section{Reinforcement learning approach to quantum control}

\subsection{Markov decision process \label{MDP section}}

We begin by introducing several concepts from the field of artificial intelligence (AI). An intelligent {\it agent} is any device that can be viewed as perceiving its {\it environment} through sensors and acting upon that environment with actuators \cite{Russel}. 
In reinforcement learning (RL) \cite{Sutton2017, Francois-Lavet2018}, a sub-field of AI, the interaction of the agent with its environment is usually described with a powerful framework of Markov decision processes (MDP).

In the MDP framework, the agent-environment interaction proceeds in {\it episodes} consisting of a sequence of discrete {\it time-steps}. At every time-step $t$ the agent receives an {\it observation}  $\,o_{t}\in{\cal O}$ containing some information about the current environment {\it state} $s_t\in{\cal S}$, and acts on the environment with an {\it action} $a_{t}\in{\cal A}$. This action induces a transition of the environment to a new state $s_{t+1}$ according to a Markov transition function ${\cal T}(s_{t+1}|s_{t},a_{t})$. The agent selects actions according to a {\it policy} $\pi(a_t|h_t)$, which in general can depend on the {\it history} $h_t=o_{0:t}$ of all past observations made in the current episode. In the {\it partially observable} environment, observations are issued according to an {\it observation function} $O(o_{t}|s_{t})$ and carry only a limited information about the state. In the special case of a {\it fully-observable} environment, the observation $o_t = s_t$ is a sufficient statistic of the past, which allows to restrict the policy to a mapping from states to actions $\pi(a_t|s_t)$. Environments can be further categorized as {\it discrete} or {\it continuous} according to the structure of the state space $\cal S$, and as {\it deterministic} or {\it stochastic} according to the structure of the transition function $\cal T$. Likewise, policies can be categorized as discrete or continuous according to the structure of the actions space $\cal A$, and as deterministic or stochastic.

The agent is guided through the learning process by a {\it reward} signal $r_{t}\in{\cal R}$. The reward is issued to the agent after each action, but it cannot be used by the agent to decide on the next action. Instead, it is used by the {\it learning algorithm} to improve the policy. 
The reward signal is designed by a human supervisor according to the final goal, and it must indicate how good the new environment state is after the applied action. 
Importantly, it is possible to specify the reward signal for achieving a final goal without knowing what the optimal actions are, which is a major difference between reinforcement learning and more widely appreciated supervised learning. 
The goal of the learning algorithm is to find a policy $\pi$ that maximizes the agent's {\it utility function} $J$, which in RL is taken to be the expectation $J=\mathbb{E}_\pi[R]$ of the reward accumulated during the episode, also known as a {\it return} $R=\sum_{t}r_{t}$.

Even from this brief description it is clear that learning environments vary vastly in complexity from ``simple'' discrete fully-observable deterministic environments, such as a Rubik's cube, to ``difficult'' continuous partially-observable stochastic environments, such as those of self-driving cars. Where does quantum control land on this spectrum?

\begin{figure}[t]
 \includegraphics[width = \figwidth]{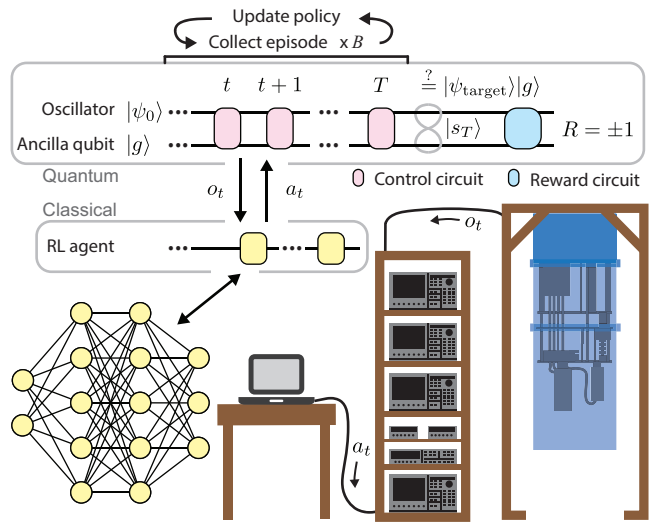}
 \caption{\label{fig1} The pipeline of classical reinforcement learning applied to a quantum-observable environment. The agent (yellow box), whose policy is represented with a neural network, is a program implemented in a classical computer controlling the quantum system. The quantum environment of the agent consists of a harmonic oscillator and its ancilla qubit, implemented with superconducting circuits and cryogenically cooled in the dilution refrigerator. The goal of the agent is to prepare the target state $|\psi_{\rm target}\rangle$ of the oscillator after $T$ time-steps, starting from initial state $|\psi_0\rangle$. Importantly, the agent does not have access to the quantum state of the environment, it can only observe the environment through intermediate projective measurements of the ancilla qubit yielding binary outcomes $o_t$. The agent controls the environment by producing at each time-step the action-vector $a_{t}$ of parameters of the control circuit (pink box). The reward $R$ for the RL training is obtained by executing the reward circuit (blue box) on the final state $|s_T\rangle$ prepared in each episode. This circuit is designed to probabilistically answer the question ``Is the prepared state $|s_T\rangle$ equal to $|\psi_{\rm target}\rangle |g\rangle$?'' A batch of $B$ episodes is collected per training epoch and used in the classical optimization loop to update the policy.}
\end{figure}

\subsection{Quantum control as quantum-observable Markov decision process \label{QOMDP}}
To explain how quantum control can be viewed as a sequential decision problem, for concreteness we specialize the discussion to a typical circuit QED \cite{Blais2020} experimental setup, depicted in Fig.~\ref{fig1}, although our framework is independent of the physical platform. The agent is a program implemented in a classical computer controlling the quantum system. The quantum environment of the agent consists of a quantum harmonic oscillator, realized as an electromagnetic mode of the superconducting resonator, and an ancilla qubit, realized as the two lowest energy levels of a  transmon \cite{Koch2007}. Note the difference in the use of the term “environment” which in quantum physics refers to dissipative bath coupled to a quantum system, while in our RL context it refers to the quantum system itself, which is the environment of the agent.

It is convenient to abstract away the exact details of the control hardware and adopt the circuit model of quantum control. According to such operational definition, the agent interacts with the environment by executing a parameterized control circuit in discrete steps, as illustrated in Fig.~\ref{fig1}. On each step $t$, the agent receives an observation $o_t$, and produces the action-vector $a_{t}$ of parameters of the control circuit to apply in the next time step. The agent-environment interaction proceeds for $T$ steps, comprising an episode. 

Compared to the typical classical partially-observable MDPs (POMDPs), there are two significant complications in the quantum case: (i) the quantum environment is minimally observable to the agent through projective ancilla measurements, i.e. the observations $o_{t}$ carry no more than $1$ bit of information, and (ii) the observation causes a random discontinuous jump in the underlying environment state. 
While in principle classical POMDPs could have such properties, they arise more naturally in the quantum case.
Historically, RL was benchmarked sometimes in stochastic but always richly observable environments, and it is therefore an open question whether existing RL algorithms are well suited for quantum environments with properties (i)-(ii). There is also a fundamental question of whether classical agents can efficiently, in the algorithmic complexity sense, learn compressed representations of the latent quantum states producing the observations, and if such representations are necessary for learning quantum control policies. Recognizing some of these difficulties, Ref.~\cite{Barry2014} introduced ``Quantum-Observable Markov Decision Process'' (QOMDP), a term we will adopt to describe our quantum control framework.

We use the Monte Carlo wave-function method \cite{Molmer1993} to simulate the quantum environment of the agent. For the environment consisting of an oscillator coupled to ancilla qubit and isolated from the dissipative bath, the most general QOMDP has the following specifications:

{\bf 1.} State space is the joint Hilbert space of the oscillator-qubit system, which in our simulation corresponds to ${\cal S}=\{|s\rangle\in  \mathbb{C}^{2}\otimes\mathbb{C}^{N}, \, \langle s|s\rangle=1 \}$, with $N=100$ being oscillator Hilbert space truncation in the photon number basis.

{\bf 2.} Observation space ${\cal O}=\{-1,+1\}$ is a set of possible measurement outcomes of the qubit $\sigma_{z}$ operator. 
If the control circuit contains a qubit measurement, the observation function is given by the Born rule of probabilities. If the control circuit does not contain a measurement, the observation is a constant which we take to be $o_t=+1$. We refer to the former as measurement-based feedback control and the latter as unitary control.

In other approaches \cite{Chen2014, Zhang2019, Porotti2019, An2020, Haug2020, Schuff2020, Wang2019a, Fosel2018, Xu2019}, an observation is a quantum state itself $o_t=|s_t\rangle$, which is not naturally compatible with real-world experiments. It could be obtained through quantum state tomography \cite{Baum2021}, but this would result in exponential scaling of the training sample complexity with system size.

{\bf 3.} Action space ${\cal A}=\mathbb{R}^{|\cal A|}$, is the space of parameters $a$ of the control circuit. It generates the set $\{{\cal K}[{a}]\}$ of continuously parameterized Kraus maps. If the control circuit contains a qubit measurement, then each map ${\cal K}[{a}]$ consists of two Kraus operators $K_{\pm}[{a}]$ satisfying the completeness relation $K_{+}^{\dagger}[a]K_{+}[a]+K_{-}^{\dagger}[a]K_{-}[a]=I$ and corresponding to observations $\pm1$. If the control circuit does not contain a measurement, then the map consists of a single unitary operator $K_0[a]$. 

{\bf 4.} State transitions happen deterministically according to $|s_{t+1}\rangle = K_0[{a_t}]|s_t\rangle$ if the control circuit does not contain a measurement, and otherwise stochastically according to $|s_{t+1}\rangle = K_\pm[{a_t}]|s_t\rangle/\sqrt{p_\pm}$ with probabilities $p_{\pm}=\langle s_t| K_{\pm}^{\dagger}[a_t]K_{\pm}[a_t]|s_t\rangle$.

In this paper, we do not consider the coupling of a quantum system to a dissipative bath, but it can be incorporated into the QOMDP by expanding the Kraus maps to include uncontrolled quantum jumps of the state $|s_t\rangle$ induced by the bath. This would lead to more complicated dynamics, but since the quantum state and its transitions are hidden from the agent, nothing would change in the RL  framework. 

In the traditional simulation-based approach to quantum control, the model for ${\cal K}[a]$ is specified, for example through the system's Hamiltonian and Schr\"{o}dinger equation, allowing for gradient-based optimization of the cost function \cite{Khaneja2005, Caneva2011, DeFouquieres2011, Leung2017, Abdelhafez2019}.
In contrast, in our approach the Kraus map ${\cal K}[a]$ is not modeled. Instead, the experimental apparatus implements ${\cal K}[a]$ exactly. In this case, the optimization proceeds at a higher level by trial-and-error learning of the patterns in the action-reward relationship. This ensures that the learned control sequence is free of model bias. 

In practice, common contributions to model bias come from frequency- and power-dependent pulse distortions in the control lines \cite{Jerger2019,Rol2020}, higher order nonlinearities, coupling to spurious modes, etc. 
Simulation-based approaches often attempt to compensate for model bias by introducing additional terms in the cost function, such as penalties for pulse power and bandwidth, weighted with somewhat arbitrarily chosen coefficients, or finding policies that are first-order insensitive to deviations in system parameters \cite{Propson2021}. In contrast, our RL agent will learn the relevant constraints automatically, since it optimizes the true unbiased objective incorporated into the reward. 

As shown in Fig.~\ref{fig1}, the reward in our approach is produced by following the training episode with the reward circuit. This circuit realizes a dichotomic POVM on the oscillator, whose binary outcome probabilistically indicates whether the applied action sequence implements the desired quantum operation. Since the agent’s goal is to maximize the expectation $J = \mathbb{E}[R]$, we require that in the state preparation QOMDPs the reward circuit is designed to satisfy the condition
\begin{align}
\underset{|\psi\rangle}{\rm argmax}\;\mathbb{E}[R]=|\psi_{\rm target}\rangle, \label{condition}
\end{align}
where expectation is taken with respect to the sampling noise in reward measurements when the state $|\psi\rangle|g\rangle$ is supplied at the input to the reward circuit.

In circuit QED, dichotomic POVMs are realized through unitary operation on the oscillator-qubit system followed by a projective qubit measurements in $\sigma_{z}$ basis.
Since the reward measurement in general will disrupt the quantum state, we only apply the reward circuit at the end of the episode, and use the reward $r_{t<T}=0$ at all intermediate time-steps. Hence, from now on we will omit the time-step index and refer to the reward as simply $R\equiv r_T$. 
Such delayed rewards are known to be particularly challenging for RL agents, because they need to make multiple action decisions during the episode, while the reward only informs whether the complete  sequence of actions was successful but does not provide feedback on the individual actions.

A common choice of reward $R$ in other approaches \cite{Chen2014, Bukov2018, Zhang2019, An2020, August2018, Haug2020, Kuo2021, Dalgaard2019, Niu2019, An2019, Baum2021} is the fidelity of the executed quantum operation. The fidelity oracle, often assumed freely available, would translate into time-consuming averaging in experiments involving quantum systems with high-dimensional Hilbert space, and is therefore prohibitively expensive from a practical point of view.

Clearly, quantum control is a ``difficult'' decision process according to a rough categorization outlined in Section~\ref{MDP section}. 
One may compare it to driving a car blind with a single sensor that provides binary-valued feedback instead of a rich visual picture of the surroundings.
In the following Section, we describe our approach to solving QOMDPs through policy gradient RL.

\subsection{Solving quantum control through policy gradient reinforcement learning}
The solution to a POMDP is a policy $\pi(a_{t}|h_{t})$ which assigns a probability distribution over actions to each possible history $h_{t}=o_{0:t}$ that the agent might see. In large problems, it is unfeasible to represent the policy as a lookup table, and instead it is convenient to parameterize it using a powerful function approximator such as a deep neural network \cite{Mnih,Mnih2015,Silver2016}. 
As an additional benefit, this representation allows the learning agent to generalize via parameter sharing to histories it has never encountered during training.
We will refer to such neural network policies as $\pi_{\theta}$ where $\theta$ represents the network parameters. It is advantageous to adopt recurrent network architectures, such as the Long Short-Term Memory (LSTM) \cite{Hochreiter1997}, in problems with variable-length inputs. In this work, we use neural networks with an LSTM layer and several fully connected layers.

The output of the policy network is the mean $\mu_\theta[h_t]$ and diagonal covariance $\sigma^2_\theta[h_t]$ of the multivariate Gaussian distribution from which the action $a_t$ is sampled on every time-step, as depicted in Fig.~\ref{fig1}. The stochastisity of the policy during the training ensures a balance between exploration of new actions and exploitation of the current best estimate $\mu_\theta$ of the optimal action. Typically, as training progresses, the agent learns to reduce the entropy of the stochastic policy and eventually converges to a near-deterministic policy. After the training is finished, the deterministic policy is obtained by choosing the optimal action $\mu_\theta$. 

In application to QOMDPs, such a stochastic action space exploration strategy means that every experimental run is performed with a different policy candidate which is evaluated with a binary reward measurement. Instead of spending the sample budget on increasing the evaluation accuracy for any given policy candidate through averaging, our strategy is to spend this budget on evaluating more policy candidates albeit with the minimal accuracy. Such strategy is explicitly tailored to the stochasticity and minimalistic observability of quantum environments, and is conceptually rather different from widely used model-free optimization methods that crucially rely on averaging to suppress noise in the cost function, as we further discuss in Appendix~\ref{NM}.

Policy gradient reinforcement learning \cite{Sutton2017, Francois-Lavet2018} provides a set of tools for learning the policy parameters $\theta$ guided by the reward signal. Even though the binary-valued reward $R$ is a non-differentiable random variable sampled from episodic interactions with the environment, its expectation $J$ depends on the policy parameters $\theta$ and it is therefore differentiable. The basic working principle of the policy gradient algorithms is to construct an empirical estimator $g_{k}$ of the gradient of performance measure $\nabla_{\theta}J(\pi_{\theta})|_{\theta=\theta_{k}}$ based on a batch of $B$ episodes of experience collected in the environment following the current stochastic policy $\pi_{\theta_{k}}$, and then perform a gradient ascent step on the policy parameters $\theta_{k+1}=\theta_{k}+\alpha g_{k}$, where $\alpha$ is the {\it learning rate}. This data collection and the subsequent policy update comprises a single {\it epoch} of training. 

Various policy gradient RL algorithms differ in their construction of the gradient estimator. In this work, we use the Proximal Policy Optimization algorithm (PPO) \cite{Schulman2017} whose brief summary is included in the Supplementary Material \cite{SuppMat}. PPO was developed to cure sudden performance collapses often observed when using high-dimensional neural network policies. It achieves this by discouraging large policy updates (hence ``proximal''), inspired by ideas from trust region optimization. The stability of PPO is essential in stochastic environments, motivating our choice of this algorithm for solving QOMDPs.

As described above, the learning process is a guided search in the policy space, where the guiding signal is the reward assigned to each attempted action sequence. Since in the state preparation QOMDP the goal is to approach arbitrarily close to the target state that resides in a {\it continuous} state space, it is tempting to think that the guiding signal needs to be of high resolution, i.e. assign different rewards to policies of different qualities, with reward difference being indicative of the quality difference. 
This condition is certainly satisfied by using fidelity as a reward \cite{Chen2014, Bukov2018, Zhang2019, An2020, August2018, Haug2020, Kuo2021, Dalgaard2019, Niu2019, An2019, Baum2021}. In contrast,  our reward-circuit-based approach breaks this condition but promises high experimental sample efficiency by virtue of not having to perform expensive fidelity estimation. However, it is not obvious that stochastic $\pm 1$ outcomes of the reward circuits are sufficient to navigate a continuous policy space and converge at all, not to mention reaching a high fidelity. For example, consider that for two policies with fidelities ${\cal F}_1>{\cal F}_2$, in our approach it is possible to receive the rewards  $R_1=-1<R_2=+1$ due to the measurement sampling noise, leading to the incorrect contribution to policy gradient.
By probabilistically comparing multiple policy candidates and performing small updates within the trust region, our proximal policy optimization is able to successfully cope with such a highly stochastic learning problem.

The next Section is devoted to empirically proving that our approach indeed leads to stable learning convergence, i.e. that the agent's performance gradually improves to a desired level and does not collapse or stagnate. We demonstrate this by training the agent to solve challenging state preparation instances. 

We also provide a simple introductory example illustrating the basic principles of our approach in Appendix~\ref{sec:qubit_flip_demo}.

\section{Results\label{results}}
Currently, direct pulse shaping with GRAPE (gradient ascent pulse engineering) is a dominant approach to quantum state preparation in circuit QED \cite{Heeres2017, Wang2019b, Hu2019a}. Nevertheless, a modular approach based on repetitive application of a parameterized control circuit has several advantages \cite{Eickbusch2021, Kudra2021}. Firstly, thanks to a reduced number of parameters, the modular approach is less likely to overfit and can generalize better under small environment perturbations. In addition, each gate in the module can be individually tested and calibrated. Finally, the modular approach is physically motivated and more interpretable, leading to a better understanding of the solution.

Our RL approach is compatible with any parameterized control circuit, including piece-wise constant parameterization used in the direct pulse-shaping. In this work, for concreteness, we made a particular choice of a control circuit based on the universal gate set consisting of the Selective Number-Dependent Arbitrary Phase gate ${\rm SNAP}(\varphi)$ and displacement $D(\alpha)$ \cite{Krastanov2015}:
\begin{align}
{\rm SNAP}(\varphi) & = \sum_{n=0}^\infty e^{i\varphi_n}|n\rangle \langle n|, \label{snap} \\
D(\alpha) & = \exp(\alpha a^\dagger - \alpha^* a). \label{displacement}
\end{align}

In practice, this gate set has been realized in the strong dispersive limit of circuit QED \cite{Heeres2015, Kudra2021}. Displacements $D(\alpha)$ are implemented with resonant driving of the oscillator, while the Berry phases $\varphi_n$ in the ${\rm SNAP}(\varphi)$ gate are created by driving the qubit resonantly with the $|g\rangle |n\rangle \leftrightarrow |e\rangle |n\rangle$ transition.
Recently, it was demonstrated that SNAP can be made first-order path-independent with respect to ancilla qubit decay \cite{Reinhold2020, Ma2020}. Furthermore, a linear scaling of the circuit depth $T$ with the state size $\langle n \rangle$ can be achieved for this approach \cite{Fosel2020}, while many interesting experimentally achievable states can be prepared with just $T\sim5$. Inspired by this finding, we parameterize our open-loop control circuit as $D^\dagger(\alpha)\,{\rm SNAP}(\varphi)\,D(\alpha)$, see Fig.~\ref{fig2}(a).

In the following Sections \ref{sec:Fock states}-\ref{sec:arbitrary states} our aim is to demonstrate that model-free RL is feasible, i.e. the learning converges to high-fidelity protocols in a realistic number of training episodes. To isolate the learning aspect of the problem, in Sections~\ref{sec:Fock states}-\ref{sec:arbitrary states} we use perfect gate implementations acting on the Hilbert space as intended by Eqs.~\eqref{snap}-\eqref{displacement}. However, the major power of the model-free paradigm is the ability to utilize available controls even when they do not produce the expected effect, tailoring the learned actions to the unique control imperfections present in the system. We focus on this aspect in Section~\ref{sec:imperfect SNAP} by training the agent with an imperfectly implemented SNAP. Moreover, the advantage of model-free RL compared to other model-free optimization methods is that it can efficiently solve problems requiring adaptive decision-making \cite{Silver2016, Silver2018, Mnih2015, Vinyals2019, Levine2015, Haarnoja2018b}. We leverage this advantage of RL in Section \ref{sec:imperfect SNAP} to learn adaptive measurement-based quantum feedback strategies compensating for imperfect SNAP implementation. Finally, in Appendix~\ref{gates} we demonstrate learning of gates for logical qubits encoded in an oscillator.

\subsection{Preparation of oscillator Fock states \label{sec:Fock states}}

One central question in our RL approach is how to assign a reward $R$ to the agent by performing a measurement on the prepared state $|s_T\rangle$. To satisfy Eq.~\eqref{condition}, it is sufficient to design the reward circuit in such a way that $\mathbb{E}[R]=f({\cal F})$ where $f$ is any monotonously increasing function of fidelity ${\cal F}$ to the target state. Although this is not necessary, we found it to be a  useful guiding principle. 
For example, the most efficient choice is to generate $R$ as an outcome of a measurement with POVM $\{\Omega_{\rm target},I-\Omega_{\rm target}\}$ where $\Omega_{\rm target}=|\psi\rangle\langle\psi|_{{\rm target}}$ is the target projector. This POVM maximizes the distinguishability of the target state from all other states \cite{Kliesch2021}. We will refer to such reward as target projector reward. If the measurement outcomes associated to this POVM are $\pm1$, then reward will satisfy $\mathbb{E}[R]=2{\cal F}-1$.

\begin{figure}[t]
 \includegraphics[width = \figwidth]{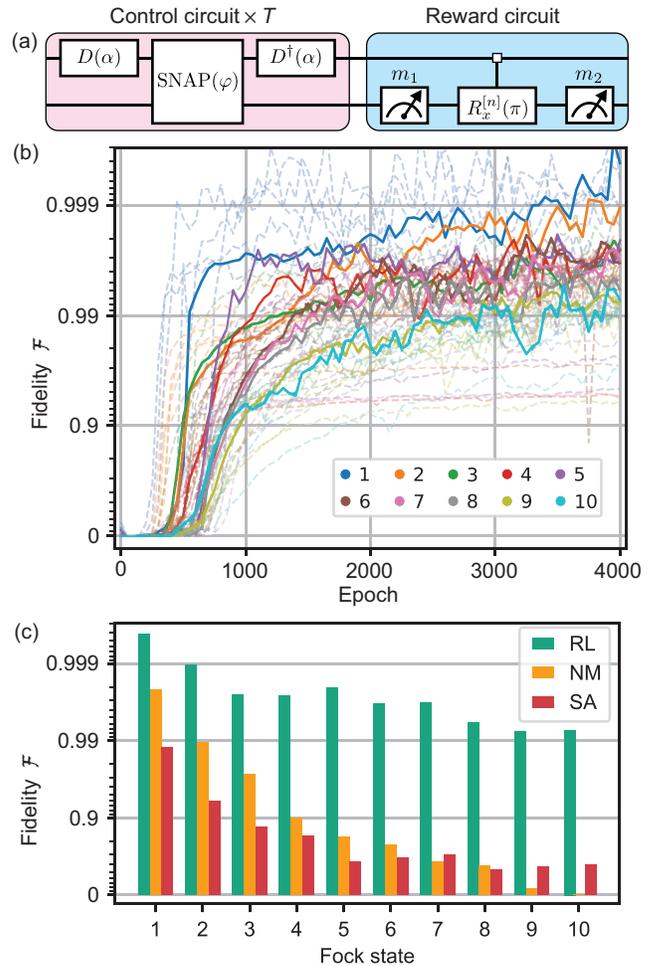}
 \caption{\label{fig2} Preparation of Fock states $|1\rangle,\, ...,\, |10\rangle$. 
 {\bf (a)} Parameterized control circuit (pink), and Fock reward circuit (blue). The  reward circuit contains a selective $\pi$-pulse on the qubit, conditioned on having $n$ photons in the oscillator. 
 {\bf (b)} Evaluation of the training progress. The background trajectories correspond to 6 random seeds for each state, solid lines show the trajectory with the highest final fidelity.
 {\bf (c)} Summary of comparison of different model-free approaches on the task of Fock state preparation. We performed extensive hyperparameter tuning for all three approaches, as described in Section~\ref{sec:Fock states} for reinforcement learning (RL), and in Appendix~\ref{NM} for Nelder-Mead (NM) and simulated annealing (SA). All approaches are constrained to the same total sample size of $M_{\rm tot}=4\cdot 10^6$. The displayed final fidelity is the highest achieved among 6 tested random seeds.  
 }
\end{figure}

In the strong dispersive limit of circuit QED \cite{Schuster2007}, a dichotomic POVM measurement required for the target projector reward can be routinely realized for an important class of non-classical states known as Fock states $|n\rangle$ that are eigenstates of the photon number operator. 
To learn preparation of such states, we use the ``Fock reward circuit'' shown in Fig.~\ref{fig2}(a).

All reward circuits considered in this work contain two ancilla measurements. If the SNAP is ideal as in Eq.~\eqref{snap}, the qubit will remain in $|g\rangle$ after the control sequence, and the outcome of the first measurement will always be $m_1=1$, which is the case in Sections~\ref{sec:Fock states}-\ref{sec:arbitrary states} and in Appendix~\ref{gates}. However, in a real experimental setup, residual entanglement between the qubit and oscillator can remain. Therefore, in general the first measurement serves to disentangle them. The second measurement with outcome $m_2$ is used to produce the reward. In the Fock reward circuit, this is done according to the rule $R=-m_2$. 

The training episodes begin with the oscillator in vacuum $|\psi_0\rangle=|0\rangle$ and the ancilla qubit in the ground state $|g\rangle$. Episodes follow the general template shown in Fig.~\ref{fig1}, in which the control circuit is applied for $T=5$ time-steps, followed by the Fock reward circuit. The SNAP gate is truncated at $\Phi=15$ levels, leading to the $(15+2)$-dimensional parameterization of the control circuit, and amounting to 85 real parameters for the full control sequence. In our approach, the choice of the circuit depth $T$ and the action space dimension $|{\cal A}|=\Phi+2$ needs to be made in advance, which requires some prior understanding of the problem complexity. In this example, we chose $T=5$ and $\Phi=15$ for all Fock states $|1\rangle$,..,$|10\rangle$ to ensure a fair comparison of the convergence speed, but, in principle, the states with lower $n$ can be prepared with shorter sequences \cite{Krastanov2015, Heeres2015}. 
An automated method for selecting the circuit depth was proposed in Ref.~\cite{Fosel2020}, and it can be utilized here to make an educated guess of $T$. 

The action-vectors are sampled from the Gaussian distribution produced by the deep neural network with one LSTM layer and two fully-connected layers, representing the stochastic policy. 
The neural network input is only the ``clock'' observation (one-hot encoding of the step index $t$), since there are no measurement outcomes in the unitary control circuit. 
The agent is trained for $4\cdot 10^3$ epochs with batches of $B=10^3$ episodes per epoch. This amounts to a sample size of $M_{\rm tot}=4\cdot 10^6$ experimental runs. The total time budget of the training is split between (i) experience collection, (ii) optimization of the neural network, and (iii) communication and instruments re-initialization. 
We estimate that with the help of active oscillator reset \cite{Pfaff2017} the experience collection time in experiment can be as short as 10 minutes in total for such training (assuming $150\, {\rm \mu s}$ duty cycle per episode). 
Our neural network is implemented with TensorFlow \cite{Abadi2016} on an NVIDIA Tesla V100 graphics processing unit (GPU). The total time spent updating the neural network parameters is 10 minutes in total for such training. 
The real experimental implementation will likely be limited by instrument re-initialization \cite{Werninghaus2021}. This time budget puts our proposal within the reach of current technology.

Throughout this manuscript, we use the fidelity $\cal F$ only as an evaluation metric to benchmark the agent, and it is not used anywhere in the training loop. If desired, in experiment the training epochs can be periodically interleaved with evaluation epochs to perform  fidelity estimation \cite{Flammia2011, DaSilva2011} for the deterministic version of the current stochastic policy. Other metrics can also be used to monitor the training progress without interruption, such as the return and entropy of the stochastic policy.

The agent benchmarking results for this QOMDP are shown in Fig.~\ref{fig2}(b). 
They indicate that our stochastic action space exploration strategy is not only able to converge, but also yields high-fidelity solutions within a realistic number of experimental runs. The agent was able to reach ${\cal F}>0.99$ for all Fock states, and ${\cal F}>0.999$ for Fock state $|1\rangle$. 

Such stable convergence in a stochastic setting is possible with proximal policy optimization because after every epoch the policy distribution only changes by a small amount within a trust region. This working principle is in stark contrast with popular optimization algorithms such as the Nelder-Mead (NM) simplex search \cite{Kelly2014, Chen2016, Rol2017} or simulated annealing (SA) \cite{Baum2021}, where each update of the simplex (in NM) or the state (in SA) can result in a drastically different policy. As a result, both these approaches perform poorly on high-dimensional problems with stochastic cost function, as shown in Appendix~\ref{NM} and summarized in Fig.~\ref{fig2}(c). When allowed the same total number of experimental runs $M_{\rm tot}=4\cdot 10^6$ as in Fig.~\ref{fig2}(b), NM is only able to find solutions with ${\cal F}>0.99$ for Fock states $|1\rangle$ and $|2\rangle$ and SA only for Fock state $|1\rangle$.

Despite its low resolution, target projector reward represents the most informative POVM from the perspective of state certification \cite{Kliesch2021}, and results in efficient learning of state preparation protocols. However, for most target states it will be unfeasible to experimentally implement such POVM in a trustworthy way. Recall that in circuit QED any dichotomic POVM on the oscillator is implemented with a unitary operation on the oscillator-qubit system and a subsequent qubit measurement in $\sigma_{z}$ basis. The trustworthiness requirement implies that this unitary operation  can be independently calibrated to high accuracy, because errors in its implementation can bias the reward circuit and, as a result, bias the learning objective of the agent. For example, in the Fock reward circuit in Fig.~\ref{fig2}(a) the unitary is a simple photon-number-selective qubit flip whose calibration is relatively straightforward.
Therefore, we consider Fock reward as a {\sl feasible and trustworthy} instance of the target projector reward.

In a more general case, when a target projector reward is unfeasible to implement, consider the following probabilistic measurement strategy. Let $\{\Omega_{k}\}$ be a parameterized set of POVM elements that can be realized in a trustworthy way. To implement a reward measurement, in each episode we first sample the parameter $k$ from some probability distribution $P(k)$, and then implement a dichotomic POVM $\{\Omega_{k},I-\Omega_{k}\}$ with associated reward $R=\pm R_k$. One can view such a reward scheme as probabilistically testing different properties of the prepared state, instead of testing directly whether it is equal to the target state. The scale $R_k$ of the binary reward is chosen according to the importance of each such property. Note that in such reward scheme the expectation in Eq.~\eqref{condition} is taken with respect to both the sampling of POVMs and the shot noise in measurement outcomes. 

In the following Sections~\ref{sec:Stabilizer states}-\ref{sec:arbitrary states}, we consider examples of such probabilistic reward measurement schemes, with further examples relevant for other physical systems included in Appendix~\ref{sec:other systems}.

\subsection{Preparation of stabilizer states \label{sec:Stabilizer states}}

The class of stabilizer states is of particular interest for quantum error correction \cite{Nielsen2010}. A state is a stabilizer state if it is a unique joint eigenvalue-1 eigenstate of a commutative stabilizer group.
To demonstrate learning stabilizer state preparation in an oscillator, we train the agent to prepare a grid state, also known as the Gottesman-Kitaev-Preskill (GKP) state \cite{Gottesman2001}. Grid states were originally introduced for encoding a 2D qubit subspace into an infinite-dimensional Hilbert space of an oscillator for bosonic quantum error correction, and were subsequently recognized to be valuable resources for various other quantum applications. In particular, the 1D version of the grid state which we consider here, can be used for sensing both real and imaginary parts of a displacement simultaneously \cite{Duivenvoorden2017, Noh2020}. 

\begin{figure}[t]
 \includegraphics[width = \figwidth]{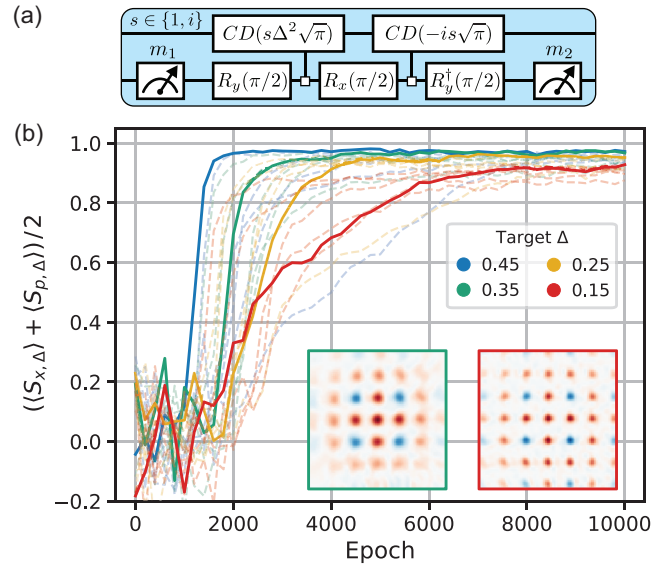}
 \caption{\label{fig3} Preparation of grid states. 
 {\bf (a)} Stabilizer reward circuit for the target state $|\psi_\Delta^{\rm GKP}\rangle$. The circuit makes use of the conditional displacement gate $CD(\alpha)=D(\sigma_z\alpha/2)$. The control circuit is the same as in Fig.~\ref{fig2}(a).
 {\bf (b)} Evaluation of the training progress. The background trajectories correspond to 6 random seeds for each state, solid lines show the trajectory with the highest final stabilizer value. 
 Inset: example Wigner functions of the states prepared by the agent after 10,000 epochs of training.}
\end{figure}

An infinite-energy 1D grid state is a Dirac comb $|\psi^{\rm GKP}_0\rangle \propto \sum_{t\in \mathbb{Z}} D(t\sqrt{\pi})|0_x\rangle$, where $|0_x\rangle$ is a position eigenstate located at $x=0$.  
The generators of a stabilizer group for such a state are $S_{x,0}=D(\sqrt{\pi})$ and $S_{p,0}=D(i\sqrt{\pi})$. The finite-energy version of this state $|\psi^{\rm GKP}_\Delta\rangle$ can be obtained with generators $S_{x,\Delta}=E_\Delta S_{x, 0} E_\Delta^{-1}$ and $S_{p,\Delta}=E_\Delta S_{p, 0} E_\Delta^{-1}$, where $E_\Delta=\exp(-\Delta^2 a^\dagger a)$ is the envelope operator, and $\Delta$ determines the degree of squeezing in the peaks of the Dirac comb and the extent of the grid envelope.

To learn preparation of such a GKP state, consider a probabilistic reward measurement scheme based on a set $\{\Omega_{k}\}$ with $k=x,p$ of POVM elements which are the projectors onto the $+1$ eigenspaces of stabilizer generators $S_{x/p,\Delta}$. The direction of the stabilizer displacement (along $x$ or $p$ quadrature) is sampled uniformly, and the scale of reward is $R_k=1$ for each direction. In this scheme, there is no simple relation between $\mathbb{E}[R]$ and ${\cal F}$, but the condition \eqref{condition} is satisfied. In contrast, for a multi-qubit system with a finite stabilizer group it is possible to construct a scheme in which the expectation of reward is a monotonous function of fidelity by sampling uniformly from the full stabilizer group, see Appendix~\ref{sec:other systems}.

The infinite-energy stabilizers $S_{x/p,0}$ are unitary and can be measured in the oscillator-qubit system with the standard phase estimation circuit \cite{Terhal2016}, as was experimentally demonstrated with trapped ions \cite{Fluhmann2019} and superconducting circuits \cite{Campagne-Ibarcq2020}. On the other hand, the finite-energy stabilizers $S_{x/p,\Delta}$ are not unitary nor Hermitian. Recently, an approximate circuit for generalized measurement of $S_{x/p,\Delta}$ was proposed \cite{Royer2020,DeNeeve2020} and realized with trapped ions \cite{DeNeeve2020}. Our stabilizer reward circuit, shown in Fig.~\ref{fig3}(a), is based on these proposals. The measurement outcome $m_2$, obtained in this circuit, is administered as a reward $R=m_2$. Since this circuit only approximates the desired POVM, such reward will only approximately satisfy $\mathbb{E}[R]=(\langle S_{x,\Delta}\rangle + \langle S_{p,\Delta}\rangle)/2$ and fullfill the condition \eqref{condition}. Nevertheless, the agent that strives to maximize such a reward will learn to prepare an approximate $|\psi^{\rm GKP}_\Delta\rangle$ state. 

After choosing the reward circuit, we need to properly constrain the control circuit. Grid states have a large photon number variance $\sqrt{{\rm var}(n)}\approx \langle n\rangle \approx 1/(2\Delta^2)$, hence preparation of such states requires a large SNAP truncation $\Phi$. However, increasing the  action space dimension $|{\cal A}|=\Phi+2$ can result in less stable and efficient learning. As a compromise, we choose $\Phi=30$ and $T=9$, amounting to  288 real parameters for the full control sequence.

The agent benchmarking results for this QOMDP are shown in Fig.~\ref{fig3}(b), with average stabilizer value as the evaluation metric [measured with the approximate circuit from Fig.~\ref{fig3}(a)]. For a perfect policy, the stabilizers would saturate to $+1$, but it is increasingly difficult to satisfy this requirement for target states with smaller $\Delta$ due to a limited SNAP truncation and circuit depth.  Nevertheless, our agent successfully copes with this task.
Example Wigner functions of the states prepared by the agent after 10,000 epochs of training are shown as insets.

Learning state preparation with a probabilistic reward measurement scheme is generally less efficient than with target projector reward because individual reward bits carry only partial information about the state. However, in principle, if stabilizer measurements can be realized in a quantum non-demolition way, this opens a possibility of acquiring the values of multiple commuting stabilizers after every episode, and thereby increasing the signal-to-noise ratio (SNR) of the reward signal.

Reward circuits in Sections~\ref{sec:Fock states}-\ref{sec:Stabilizer states} are designed for special classes of states. Next, we consider how to construct a reward circuit applicable to arbitrary states.

\subsection{Preparation of arbitrary states \label{sec:arbitrary states}}

In the general case, we aim to construct an unbiased estimator of fidelity $\cal F$ based on a measurement scheme which is (i)  tomographically complete, (ii) feasible to implement in a given experimental platform, and (iii) trustworthy. The requirement (i) in combination with universality of the control circuit is necessary to guarantee that arbitrary states can in principle be prepared with our approach. However, it is not by itself sufficient, and needs to be supplemented with requirements (ii) and (iii) to ensure practical feasibility.

In the strong dispersive limit of circuit QED, the Wigner tomography is a canonical example satisfying all three requirements above \cite{Vlastakis2013}. Wigner function is defined on the oscillator phase space with coordinates $\alpha\in\mathbb{C}$, and is given as the expectation value of the ``displaced parity'' operator $W(\alpha)=\frac{2}{\pi}\langle \Pi_\alpha\rangle$, where $\Pi_\alpha=D(\alpha) \Pi D^\dagger(\alpha)$, and $\Pi=e^{i\pi a^\dagger a}$ is the photon number parity. Hence, for the probabilistic reward measurement scheme based on the Wigner function, we consider a continuously parameterized set of POVM elements $\{\Omega_\alpha\}$, where $\Omega_\alpha = (I + \Pi_\alpha)/2$ is a projector onto +1 (even) eigenspace of the displaced parity operator.

\begin{figure}[t]
\includegraphics[width = \figwidth]{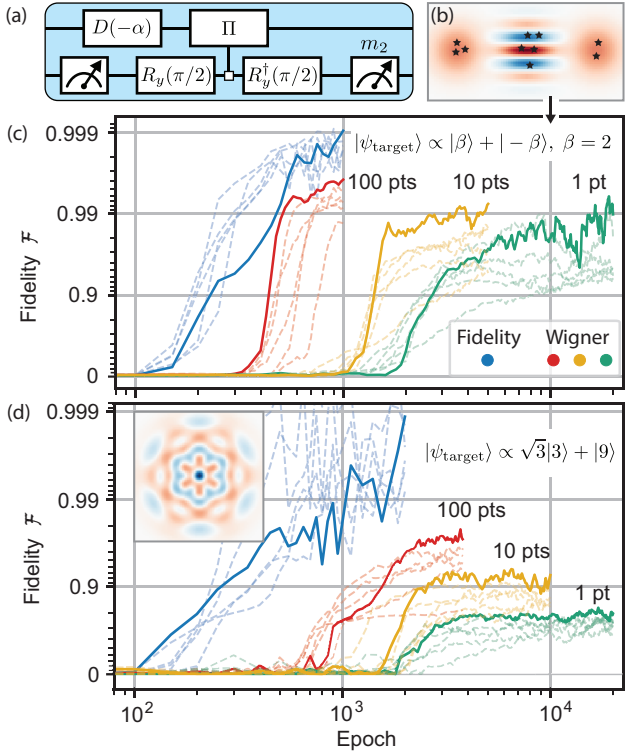}
\caption{\label{fig4} Preparation of arbitrary states. 
{\bf (a)} Wigner reward circuit based on the measurement of the photon number parity. In this circuit, the conditional parity gate corresponds to $|g\rangle\langle g|\otimes I + |e\rangle \langle e|\otimes \Pi$.
{\bf (b)} Wigner function of the cat state $|\psi_{\rm target}\rangle \propto|\beta\rangle + |-\beta\rangle$ with $\beta=2$. Scattered stars illustrate phase space sampling of points $\alpha$ for the Wigner reward. 
{\bf (c)} Evaluation of the training progress for the cat state. The background trajectories correspond to 6 random seeds for each setting, solid lines show the trajectory with the highest final fidelity. The Wigner reward is obtained by sampling $1,10,100$ different phase space points, doing a single measurement per point and averaging the obtained measurement outcomes to improve the resolution and achieve higher convergence ceiling. For blue curves the fidelity $\cal F$ is used as a reward, representing the expected performance in the limit of infinite averaging.
{\bf (d)} Evaluation of the training progress for the binomial code state $|\psi_{\rm target}\rangle \propto \sqrt{3}|3\rangle + |9\rangle$, whose Wigner function is shown in the inset.}
\end{figure}

Next, we need to determine the probability distribution $P(\alpha)$ according to which the POVMs are samples from the set $\{\Omega_\alpha\}$ for reward evaluation. To this end, we derive the estimator of fidelity based on the Monte Carlo importance sampling of the phase space:
\begin{align}
{\cal F} & = \pi\int d^{2}\alpha\,W(\alpha)W_{{\rm target}}(\alpha) \label{integral} \\
&=2 \underset{\alpha\sim P}{\mathbb{E}}\, \underset{\psi}{\mathbb{E}} \left[ \frac{1}{P(\alpha)} \Pi_\alpha W_{{\rm target}}(\alpha)\right],
\label{fidelity_estimator_wigner}
\end{align}
where points $\alpha$ are sampled according to an arbitrary probability distribution $P(\alpha)$ which is nonzero where $W_{{\rm target}}(\alpha)\neq0$. 
The estimator \eqref{fidelity_estimator_wigner} leads to the following scheme, dubbed ``Wigner reward'': first, the phase space point $\alpha$ is generated with rejection sampling, as illustrated in Fig.~\ref{fig4}(b), and then the displaced parity $\Pi_\alpha$ is measured, corresponding to the reward circuit shown in Fig.~\ref{fig4}(a). Reward is then assigned according to the rule $R = R_\alpha m_2$, where  $R_\alpha=\frac{2c}{P(\alpha)}W_{{\rm target}}(\alpha)$ is chosen to reflect the importance of a sampled phase space point, and $c>0$ is an arbitrary scaling factor. Such reward satisfies $\mathbb{E}[R] = c\,{\cal F}$ according to Eq.~\eqref{fidelity_estimator_wigner}, but only requires a single binary tomography measurement per policy candidate.

The estimator \eqref{fidelity_estimator_wigner} is unbiased for any $P(\alpha)$, but its variance can be reduced by choosing $P(\alpha)$ optimally. The lowest variance is achieved with $P(\alpha)\propto |W_{{\rm target}}(\alpha)|$, as shown in Appendix~\ref{sec: variance}. Such a choice also helps to stabilize the learning algorithm, since it conveniently leads to rewards $R=m_2\, {\rm sgn}\,W_{{\rm target}}(\alpha)$ of equal magnitude $|R|=1$, where we made a proper choice of the scaling factor $c$. 

We investigate the agent's performance with Wigner reward circuit for (i) preparation of the Schr{\"o}dinger cat state $|\psi_{\rm target}\rangle\propto|\beta\rangle + |-\beta\rangle$ with $\beta=2$ in $T=5$ steps, shown in Fig.~\ref{fig4}(c), and (ii) preparation of the binomial code state $|\psi_{\rm target}\rangle \propto \sqrt{3}|3\rangle + |9\rangle$ \cite{Michael2016} in $T=8$ steps, shown in Fig.~\ref{fig4}(d). In contrast to target projector and stabilizer rewards that asymptotically lead to reward of $+1$ for optimal policy, Wigner reward remains stochastic even under the optimal policy. Since in this case it is impossible to find the policy that would systematically produce a reward of $+1$, for some states the agent converges to policies of intermediate fidelity (green). To increase the SNR of the Wigner reward, we evaluate every policy candidate with reward circuits corresponding to $1,10,100$ different phase space points, doing a single measurement per point and averaging the obtained measurement outcomes to generate the reward $R$.  The results show that increased reward SNR allows to reach higher fidelity, albeit at the expense of increased sample size. We expect that in the limit of infinite averaging the training would proceed as if the fidelity $\cal F$ was directly available to be used as reward (blue).

We observe notable variations in convergence speed and saturation fidelity depending on the choice of hyperparameters, which is typical of reinforcement learning. A lot of progress has been made in developing robust RL algorithms applicable to a variety of tasks without extensive problem-specific hyperparameter tuning \cite{Mnih2015, Silver2018}, but this still remains a major open problem in the field. The list of hyperparameters used in all our training examples can be found in the Supplementary Material  \cite{SuppMat}. Even with the optimal choice of hyperparameters, there is no rigorous guarantee of convergence -- a problem plaguing all heuristic optimization methods in non-convex spaces. In the presented examples, we plot learning trajectories corresponding to several random seeds to demonstrate that the probability of getting stuck with a suboptimal solution is small.

This demonstration shows that arbitrary state preparation is in principle possible with our approach, as long as a tomographically complete reward measurement scheme is available in a given physical system. In Appendix~\ref{sec:other systems}, we provide fidelity estimators  based on the characteristic function, enabling training for arbitrary state preparation in trapped ions and multi-qubit systems.

\begin{figure*}[t]
 \includegraphics[width = \figwidthDouble]{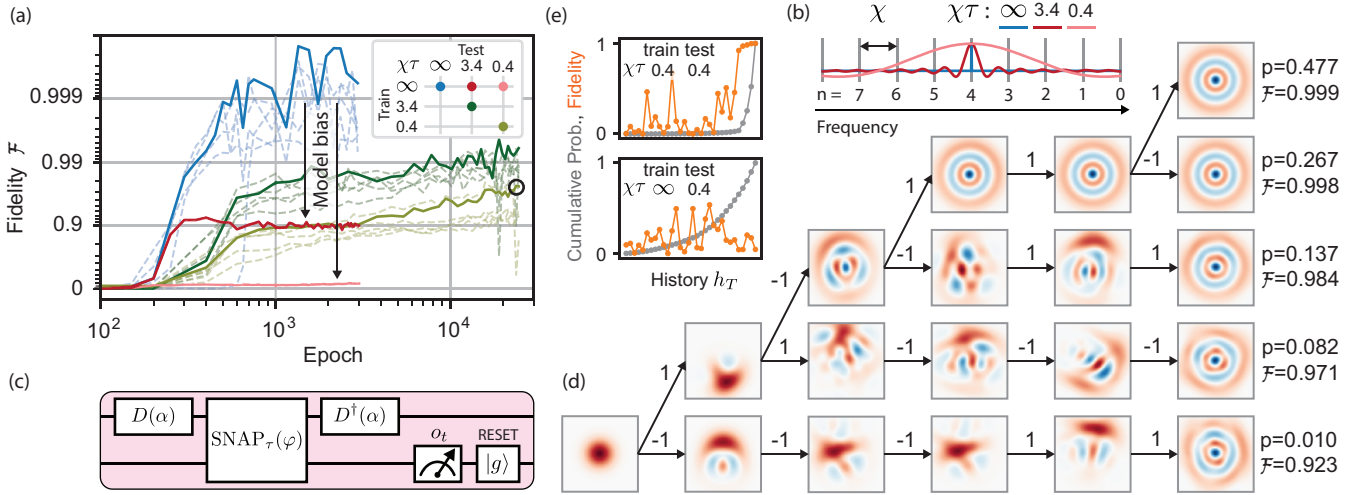}
 \caption{\label{fig5} Learning adaptive measurement-based quantum feedback for preparation of Fock state $|3\rangle$ with imperfect controls.
{\bf (a)} Evaluation of the training progress. Blue: training the agent with the open-loop control circuit, shown in Fig.~\ref{fig2}(a), that uses an ideal SNAP. The background trajectories correspond to 6 random seeds. The protocols of the best-performing seed are then tested using the same control circuit, but with a finite-duration gate ${\rm SNAP}_\tau$ substituted instead of an ideal SNAP. Such a test reveals the degradation of performance (red, pink) due to the model bias. 
{\bf (b)} Spectrum of partially-selective qubit pulses used in the gate ${\rm SNAP}_\tau$. The degradation of performance in (a) occurs because the pulse overlaps in the frequency domain with unintended number-split qubit transitions, leaving the qubit and oscillator entangled after the gate. 
{\bf (c)} Closed-loop control circuit containing a finite-duration gate ${\rm SNAP}_\tau$ and a verification measurement that produces an observation $o_t$ and disentangles qubit and oscillator. The qubit is always reset to $|g\rangle$ after the measurement. This control circuit requires either post-selection or adaptive control. The agent successfully learns measurement-based feedback control (a, green) even in the extreme case $\chi\tau=0.4$ far from theoretically optimal regime $\chi\tau\gg1$. 
{\bf (d)} An example state evolution under the policy obtained after 25,000 epochs of training, shown with a black circle in (a). The agent chooses to focus on a small number of branches and ensure that they lead to high-fidelity states.
{\bf (e)} Cumulative probability and fidelity of the observed histories quantifies this trend (top panel). The policy trained with ideal SNAP and tested with ${\rm SNAP}_{\tau}$ (bottom panel) has relatively uniform probability of all histories and poor fidelity.}
\end{figure*}

Examples considered in Sections~\ref{sec:Fock states}-\ref{sec:arbitrary states} already demonstrate the model-free aspect of our approach despite the  perfect gate implementations in the underlying simulation of the quantum state evolution. In the following example, we demonstrate this aspect more explicitly by training the agent on a system with imperfect SNAP. In addition, the next example highlights the potential of RL for measurement-based feedback control. 

\subsection{Learning adaptive quantum feedback with imperfect controls \label{sec:imperfect SNAP}}

Many quantum control experiments with circuit QED systems claim decoherence-limited fidelity \cite{Heeres2015,Heeres2017}. The  effect of decoherence on the quantum operation can be decreased by reducing the execution time. However, this would involve controls with wider spectrum and larger amplitude, pushing the system to the limits where model assumptions are no longer valid. Therefore, such experiments are decoherence-limited instead of model-bias-limited only {\sl by choice}. Recent experiments that push quantum control towards faster implementation \cite{Eickbusch2021, Kudra2021} reveal that significant parts of the error budget cannot be accounted for by common and well-understood theoretical models, making the problem of model bias explicit. Model-free optimization will become an indispensable tool to achieve higher experimental fidelity despite the inability to capture the full complexity of a quantum system with a simple model.

To provide an example of this effect, we consider again a SNAP-displacement control sequence. In the oscillator-qubit system with dispersive coupling $H_c/h=\frac{1}{2}\chi\, a^\dagger a\,\sigma_z$, the Berry phases $\varphi_n$ in \eqref{snap} are created through photon-number-selective qubit rotations:
\begin{align}
{\rm SNAP}(\varphi) = \sum_n |n\rangle \langle n| \otimes R_{\pi-\varphi_n}(\pi)R_0(\pi),
\end{align}
where $R_\phi(\vartheta)=\exp(-i\frac{\vartheta}{2}[\cos\phi\, \sigma_x+\sin\phi\, \sigma_y])$. 
Note that this operation, if implemented perfectly, would return the qubit to the ground state, and hence it can be considered as an operation on the oscillator alone, as defined in \eqref{snap}.
Such an implementation relies on the ability to selectively address number-split qubit transitions, which requires pulses of long duration $\tau\gg1/\chi$. In practice, it is desirable to keep the pulses short to reduce the probability of ancilla relaxation during the gate. However, shorter pulses of wider bandwidth would drive unintended transitions, as illustrated in Fig.~\ref{fig5}(b), leading to imperfect implementation of the SNAP gate: in addition to accumulating incorrect Berry phases for different levels, this will generally leave the qubit and oscillator entangled. Such imperfections are notoriously difficult to calibrate out or precisely account for at the pulse or sequence construction level, which presents a good testbed for our model-free learning paradigm. 
We demonstrate that our approach leads to high-fidelity protocols even in the case $\tau<1/\chi$ far from theoretically optimal regime, where the sequences produced assuming ideal SNAP yield poor fidelity due to severe model bias.

We begin by illustrating in Fig.~\ref{fig5}(a) the degradation of performance of the policies optimized for preparation of Fock state $|3\rangle$ using the open-loop control circuit from Fig.~\ref{fig2}(a) with an ideal SNAP (blue), when tested with a finite-duration gate ${\rm SNAP}_\tau$ (red, pink) whose details are included in the Supplementary Material \cite{SuppMat}. Achieving extremely high fidelity (blue) requires delicate adjustment of the control parameters, but this fine-tuning is futile when the remaining infidelity is smaller than the performance gap due to model bias, shown with arrows in Fig.~\ref{fig2}(a) and a priori unknown.
As seen by testing on the $\chi\tau=3.4$ case (red), any progress that the optimizer made after 300 epochs was due to overfitting to the model of the ideal SNAP. As depicted with a spectrum in Fig.~\ref{fig5}(b), the qubit pulse of such duration is still reasonably selective (and is close to the experimental choice $\chi\tau\approx 4$ in \cite{Heeres2015}), but it already requires a much more sophisticated modeling of the SNAP implementation in order to not limit the experimental performance. In the partially selective case $\chi\tau=0.4$ (pink) the performance is drastically worse. Note that sequences optimized with any other simulation-based approach assuming ideal SNAP, such as \cite{Krastanov2015,Fosel2020}, would exhibit a similar degradation. 

One way to recover higher fidelity is through a detailed modeling of the composite qubit pulse in the SNAP \cite{Kudra2021}, although such approach will still contain residual model bias. An alternative approach, which comes at the expense of reduced success rate, is to perform a verification ancilla measurement and post-selection, leading to a control circuit shown in Fig.~\ref{fig5}(c). Post-selecting on a qubit measured in $|g\rangle$ in all time steps (history $h_T=11111$) significantly boosts the fidelity of a biased policy from $0.9$ to $0.97$ in the case $\chi\tau=3.4$, but does not lead to any improvement in the extreme case $\chi\tau=0.4$. The post-selected fidelity is still lower than with the ideal SNAP, because such a scheme only compensates for qubit under- or over-rotation, and not for the incorrect Berry phases. Additionally, the trajectories corresponding to other measurement histories have extremely poor fidelities because only the history $h_T=11111$ was observed during the optimization with an ideal SNAP. 

However, in principle, if the qubit is projected to $|e\rangle$ by the measurement, the desired state evolution can still be recovered using adaptive quantum feedback. 
Experimental Fock state preparation with quantum feedback was demonstrated in the pioneering work in cavity QED \cite{Sayrin2011}.
In our context, a general policy in the adaptive setting is a binary decision tree, equivalent to $2^{T-1}$ distinct parameter settings for every possible measurement history. There exist model-based methods for construction of such a tree \cite{Shen2017}, but they are not applicable in the cases dominated by a-priori unknown control errors. An RL agent, on the other hand, can discover such a tree in a model-free way. Even though our policies are represented with neural networks, they can be easily converted to a  decision tree representation which is more advantageous for low-latency inference in real-world experimental implementation. 

To this end, we train a new agent with a feedback-based control circuit that directly incorporates a finite-duration imperfect gate ${\rm SNAP}_\tau$, shown in Fig.~\ref{fig5}(c), mimicking training in an experiment. We use Fock reward circuit, shown in Fig.~\ref{fig2}(a), in which $m_1=1$ in all episodes despite the imperfect SNAP because of the qubit reset operation.
Since the control circuit contains a measurement, the agent will be able to dynamically adapt its actions during the episode depending on the received outcomes $o_t$.
As shown with the green curves in Fig.~\ref{fig5}(a), the agent successfully learns adaptive strategies of high fidelity even in the extreme case $\chi\tau=0.4$. This indicates that RL is not only good for fine-tuning or ``last-mile'' optimization, but is a valuable tool for the domains where model-based quantum control is not applicable, e.g. due to absence of reliable models or prohibitive memory requirements for simulation of a large Hilbert space.

To further analyze the agent's strategy, we select the best-performing random seed for the case $\chi\tau=0.4$ after 25,000 epochs of training and visualize the resulting state evolution in Fig.~\ref{fig5}(d). The average fidelity of such policy is ${\cal F} = 0.974$. There are 5 high-probability branches, all of which yield ${\cal F}>0.9$, and further post-selection of history $h_T=1\overline{1}111$ will boost the fidelity to ${\cal F}>0.999$. We observe that fidelity reduces in the branches with more $``\text{-}1"$ measurement outcomes (top to bottom), because, being less probable, such branches receive less attention from the agent during the training. As shown in Fig.~\ref{fig5}(e) top panel, the agent chooses to focus only on a small number of branches (5 out of $2^5$) and ensure that they lead to high-fidelity states. This is in contrast to the protocol optimized with the ideal SNAP and tested with ${\rm SNAP}_\tau$ (bottom panel), which, as a result of model bias, performs poorly and has relatively uniform probability of all histories (of course, such protocol would produce only history $11111$ if it was applied with an ideal SNAP).

It is noteworthy that in the two most probable branches in Fig.~\ref{fig5}(e) the agent actually finishes preparing the state in just $3$ steps, and in the remaining time chooses to simply idle instead of further entangling the qubit with the oscillator and subjecting itself to additional measurement uncertainty. In the other branches, this extra time is used to catch up after previously receiving undesired measurement outcomes. This indeed seems to be an intelligent strategy for such a problem, which serves as a positive indication that this agent will be able to cope with incoherent errors by shortening the effective sequence length.

We emphasize that even though for this numerical demonstration of model-free learning we had to build a specific model of the finite-duration SNAP, the agent is completely agnostic to it by construction. The only input that the agent receives is binary measurement outcomes, whose source is a black box to the agent. 
Effectively, in this demonstration the model bias comes from the mismatch between ideal and finite-duration SNAP. 
We also tested the agent against other types of model bias: we added independent random static offsets to the Berry phases and qubit rotation angles, and found that the agent performs equally well in this situation. 

\section{Discussion \label{sec:Discussion}}

As empirically demonstrated in Section~\ref{results}, our stochastic policy optimization is stable and leads to high sample efficiency.
Starting from a random initial policy, learning the preparation of high-fidelity Fock states (with target projector reward) and GKP states (with stabilizer reward) required $10^6-10^7$ experimental runs, and learning with Wigner reward required $10^7-10^8$ runs. 
Although seemingly large, this sample size compares favorably with the number of measurements required to merely tomographically verify the states of similar quality in experiments, e.g. $3\cdot 10^6$ for Fock states \cite{Heeres2017} and $2\cdot 10^7$ for GKP states \cite{Campagne-Ibarcq2020}.

Exactly quantifying the sample complexity of heuristic learning algorithms remains difficult. However, we can qualitatively establish the general trends.
A natural question to ask is whether our approach will scale favorably with increased (i) target state complexity, (ii) action space, (iii) sequence length. 

{\bf (i) Target state complexity.}  Sample efficiency of learning the control policy is affected by multiple interacting factors, but among the most important is the variance of the fidelity estimator used for the reward assignment. Variance of the estimator in Eq.~\eqref{fidelity_estimator_wigner} with $P(\alpha)\propto |W_{\rm target}(\alpha)|$ is given by  ${\rm Var} = 4(1+{\delta}_{\rm target})^2 - {\cal F}^2$, where ${\delta}_{\rm target}=\int |W_{\rm target}(\alpha)|d\alpha-1$ is one measure of the state non-classicality known as the Wigner negativity \cite{Kenfack2004} (see Appendix~\ref{sec: variance} for the derivation). This result leads to a simple lower bound on the sample complexity of learning the state preparation policy that reaches the fidelity $\cal F$ to the desired target state
\begin{align}
M > \frac{4(1+{\delta}_{\rm target})^2 - {\cal F}^2}{(1-{\cal F})^2}. \label{bouund}
\end{align}
This expression bounds the number of measurements $M$ required for resolving the fidelity $\cal F$ of a fixed policy with standard error of the mean  comparable to the infidelity.
The task of the RL agent is more complicated, since it needs to not only resolve the fidelity of the current policy, but at the same time learn how to improve it. Therefore, this bound is not tight, and the practical overhead depends on the choice of control parameterization, learning algorithm and its hyperparameters.
However, the bound \eqref{bouund} clearly indicates that learning the preparation of larger non-classical states is increasingly difficult, as one would expect, and the difficulty can be quantified according to the Wigner negativity of the state.
This is a fundamental limitation on the learning efficiency with the Wigner reward, which can only be overcome by designing a reward scheme that takes advantage of the special structure of the target state and available trustworthy state manipulation tools, as we did, for instance, for Fock states and GKP states. 
The Wigner negativity of Fock states grows as $\sqrt{n}$ \cite{Kenfack2004}, where $n$ is the photon number, which would result in $O(n)$ scaling of the bound \eqref{bouund}. In contrast, target projector reward, of which Fock reward is a special case, has target-state-independent variance ${\rm Var}={\cal F}(1-{\cal F})$ leading to a bound $M> {\cal F}/(1-{\cal F})$ which does not increase with the photon number.
How such reward design can be optimized in general is a matter that we leave for further investigation.

{\bf (ii) Action space.} The overhead on top of Eq.~\eqref{bouund} is determined, among other factors, by the choice of the control circuit. In the case of SNAP and displacement, the action space dimension $|{\cal A}|=\Phi+2$ has to grow with the target state size to ensure  individual control of the phases of involved oscillator levels. This might be problematic, since the performance of RL (or any other approach) usually declines on high-dimensional tasks, as evidenced, for instance, by studies of robotic locomotion with different numbers of controllable joints \cite{Schulman2015, Duan2016}. 
However, the sample complexity is not a simple function of $|\cal A|$, as can be inferred from Fig.~\ref{fig2}(b) where we use the same $|{\cal A}|=17$ for all Fock states. For lower Fock states, the agent quickly learns to disregard the irrelevant action dimensions because their contribution to policy gradient averages to zero. In contrast, for higher Fock states it needs to discover the pattern of relations between {\sl all} action dimensions across different time-steps, and thus the learning is slower. Note that on the same problem a much stronger degradation is observed when using the Nelder-Mead approach or simulated annealing, see Fig.~\ref{fig2}(c).

{\bf (iii) Sequence length.} Tackling decision-making problems with long-term dependencies (i.e. $T\gg1$) is what made RL popular in the first place, as exemplified by various game-playing agents \cite{Silver2016, Silver2018, Mnih2015, Vinyals2019}. In quantum control, the temporal structure of the control sequences can be exploited by adopting recurrent neural network architectures, such as the LSTM used in our work. Recently, machine learning for sequential data has significantly advanced with the invention of the Transformer models \cite{Vaswani2017} which use attention mechanisms to ensure that the gradients do not decay with the sequence depth $T$. Machine learning innovations such as this will undoubtedly find applications in quantum control. 

As can be seen above, there are some aspects of scalability that are not specific to quantum control, but are common in any control task. The generality of the model-free reinforcement learning framework makes it possible to transfer the solutions to such challenges, found in other domains, to quantum control problems.

Let us now return to the discussion of other factors influencing the sample efficiency. As we briefly alluded to previously, the overhead on top of Eq.~\eqref{bouund} depends on the learning algorithm and its hyperparameters.
Model-free RL is known to be less sample efficient than gradient-based methods, typically requiring millions of training episodes \cite{Francois-Lavet2018}. This is especially true for {\it on-policy} RL algorithms, such as PPO, since they discard the training data after each policy update. In contrast, {\it off-policy} methods keep old experiences in the replay buffer and learn from them even after the current policy has long diverged from the old policy under which the data was collected, typically resulting in better sample efficiency. Our pick of PPO was motivated by its simplicity and stability in the stochastic setting, but it is worth exploring an actively expanding collection of RL algorithms \cite{Francois-Lavet2018}, and understanding which are most suitable for quantum-observable environments.

The sample efficiency of model-free RL in the quantum control setting can be further improved by utilizing the strength of conventional simulation-based methods. A straightforward way to achieve this would be through supervised pre-training of the agent's policy in the simulation. Such pre-training would provide a better initial point for the agent subsequently re-trained in the real-world setting. Our preliminary numerical experiments show that this indeed provides significant speedups. 

The proposals discussed above resolve the bias-variance trade-off in favor of complete bias elimination, necessarily sacrificing sample efficiency. In this respect, model-free learning is a swing in the opposite direction from the traditional approach in physics of constructing sparse physically-interpretable models with very few parameters which can be calibrated in experiment. Building on the insights from machine learning community, model bias can in principle be strongly reduced (not eliminated) by learning a richly parameterized model, either physically motivated \cite{Krastanov2019a,Krastanov2020} or neural-network-based \cite{Flurin2018,Banchi2018}, from direct interaction with a quantum system. The learned model can then be used to optimize the control policy with simulation-based (not necessarily RL) methods. Another promising alternative is to use model-based reinforcement learning techniques \cite{Plaat2020}, where the agent can plan the actions by virtually interacting with its learned model of the environment while refining both the model and the policy using real-world interactions. Finally, in addition to adopting existing RL algorithms, a worthwhile direction is to design new algorithms tailored to the specifics of quantum-observable environments.

\section{Conclusion}

Addressing the problem of model bias as an inherent limitation of the dominant simulation-based approach to quantum control, we claim that end-to-end model-free reinforcement learning is not only a feasible alternative, but is also a powerful tool which will extend the capabilities of quantum control to domains where simulation-based methods are not applicable.
By focusing on control of a harmonic oscillator in the circuit QED architecture, we explored various aspects of learning under the conditions of quantum uncertainty and scarce observability. Our policy exploration strategy is explicitly tailored to these features of the quantum learning environments. We demonstrated stable learning directly from stochastic binary measurement outcomes, instead of relying on averaging to eliminate stochasticity as is done in other model-free quantum control optimization methods. With multiple numerical experiments, we confirmed that such strategy leads to high fidelity and sample efficiency on challenging control tasks that include both the unitary control and control with adaptive measurement-based quantum feedback. The RL agent that we developed can be directly applied in real-world experiments with various physical systems.

\section{Acknowledgment\label{sec:acknowledgement}}

We acknowledge a helpful discussion with Thomas F\"{o}sel. We thank the anonymous reviewers for their comments which encouraged us to make several additions.
We thank Yale Center for Research Computing for providing compute resources. 
This research is supported by ARO under Grant No. W911NF-18-1-0212. 
\appendix

\section{Educational example \label{sec:qubit_flip_demo}}

In this Appendix, we analyze a deliberately simple problem with a purpose of illustrating in detail various components and stages of the learning process.

\begin{figure}[h!]
 \includegraphics[width = \figwidth]{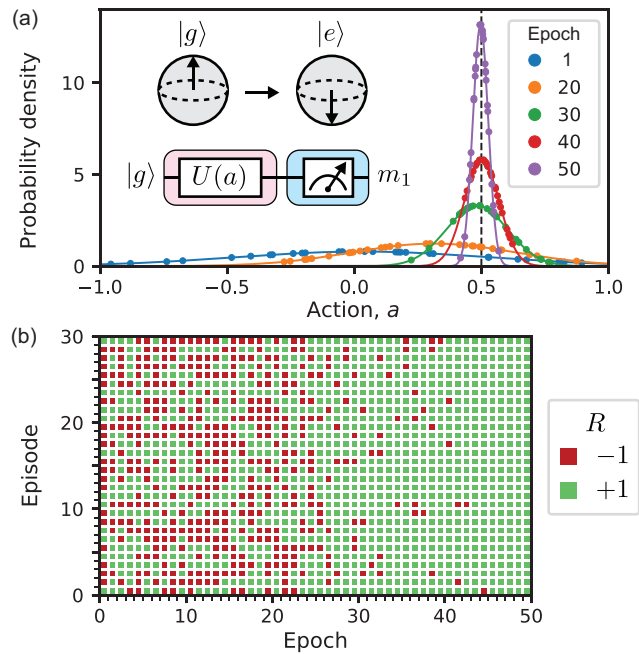}
 \caption{\label{fig6}
Educational example of model-free learning.
{\bf (a)} Inset: the task is to prepare qubit state $|e\rangle$ starting from state $|g\rangle$. Episodes consist of a single time-step; control circuit contains a rotation unitary $U(a)$, reward circuit contains a measurement of $\sigma_z$. Main panel: policy distribution (solid lines) for a selected set of epochs, and actions that the agent tried in the episodes of corresponding epoch (dots).
{\bf (b)} Rewards received by the agent in the episodes of every epoch.
}
\end{figure}

\textbf{Problem setting.} Consider a qubit state preparation problem, in which the initial state is $|g\rangle$ and the target state is $|e\rangle$. Such state preparation can be achieved with a unitary rotation gate parameterized as $U(a)=\exp(-i\pi a\sigma_{x})$, where the optimal solution $a=0.5$ is known in advance. We will let the agent discover this solution in a model-free way, without knowing which unitary is actually applied. The training episodes consist of a single time-step in which the agent produces an action $a\in\mathbb{R}^{1}$, leading to execution of control circuit $U(a)$, and then collects a reward with a simple reward circuit consisting of a $\sigma_{z}$ measurement, as show in the inset of Fig.~\ref{fig6}(a). The resulting measurement outcome $m\in\{-1,1\}$ is used to issue a reward $R=-m$, which is maximized in the target state $|e\rangle$, hence satisfying \eqref{condition}. 

\textbf{Actor and critic.} In every training episode, the action $a$ is sampled according to the probability distribution specified by the policy.
Policy $\pi_{\theta}(a)$ is parameterized with learnable parameters $\theta$.
In this problem, it is convenient to chose a simple Gaussian policy
\begin{align}
\pi_{\theta}(a)=\frac{1}{\sqrt{2\pi\sigma^{2}}}\exp\bigg[-\frac{(a-\mu)^{2}}{2\sigma^{2}}\bigg],
\label{stochastic policy}
\end{align}
whose learnable parameters are $\theta=\{\mu,\sigma^2\}$. The policy defines how the agent interacts
with the environment, and it is often referred to as {\it actor}. Another important component of PPO is the value function $V_{\theta'}$, or {\it critic}, which helps the agent asses the value of the environment state, see Supplementary Material \cite{SuppMat}. In this example, the value function can be chosen as a simple baseline $V_{\theta'}=b$ with learnable parameters $\theta'=\{b\}$. 
During the training process, parameters $\{\mu,\sigma^2,b\}$ are iteratively updated
according to the PPO algorithm.

\textbf{Training process.} The training process, illustrated in Fig.~\ref{fig6}, is split into $50$ epochs. Within each epoch $k$ the parameters of the policy remain fixed, and the agent collects
a batch of $B=30$ episodes of experience, behaving stochastically according to the current policy $\pi_{\theta_{k}}(a)$. Fig.~\ref{fig6}(a) shows the policy distribution for a selected set of epochs, and the actions that the agent tried in the episodes of corresponding epoch. The initial policy is widely distributed to ensure that the agent can adequately explore the action space. Since initially most of the actions do not lead to high-fidelity states,  the agent is very likely to receive negative rewards, as shown in Fig.~\ref{fig6}(b). After every epoch the parameters of the stochastic policy \eqref{stochastic policy} are updated
$\theta_{k}\to\theta_{k+1}$ in a way that utilizes the information contained in the reward signal. Controlled by the learning rate, these updates result in gradually shifting the probability density of the stochastic policy towards more promising actions, as seen in Fig.~\ref{fig6}(a). After iterating in this manner for several epochs, the policy becomes localized near the correct value of the action, which leads to a significantly increased fraction of positive rewards. In the initial stage the best progress is achieved by rapidly learning the parameter $\mu$. However, to achieve high fidelity it is necessary to localize $\mu$ more finely, and thus in the later stages the agent shrinks the variance
$\sigma^2$ of the policy. Eventually there are almost no episodes with negative reward, meaning that the agent has achieved good performance.

\textbf{Complications.} This simple example illustrates how learning proceeds in our approach. More realistic examples contained in Section~\ref{results} follow the same basic principles. Additional complications arise from the following considerations:

(i) Typically the action space ${\cal A}$ is high-dimensional. In such case the Gaussian policy distribution is defined on $\mathbb{R}^{|{\cal A}|}$ instead of $\mathbb{R}$.

(ii) The agent can receive a nontrivial observation $o$, for instance a qubit measurement outcome, which requires incorporating adaptive measurement-based feedback into the policy. In such case, the policy distribution $\pi_{\theta}(a|o)$ is conditioned on the observation. In case of a Gaussian policy, this is achieved by making the mean and variance be parameterized functions of the observation $\{\mu,\sigma^2\}=\{\mu_{\theta}(o),\sigma^2_{\theta}(o)\}$. In our work, these functions are chosen to be neural networks.

(iii) The episodes typically consist of multiple time-steps. In such case, the policy distribution $\pi_{\theta}(a|t;h_t)$ is conditioned on the time-step index $t$ and on the history of observations $h_t=o_{0:t}$ received up to the current time-step. For notational simplicity we usually treat the time dependence as implicit and denote the policy as $\pi_{\theta}(a|h_t)$.

\section{Alternative model-free approaches \label{NM}}

\subsection{Qualitative comparison of action space exploration strategies}

\begin{figure}[t]
 \includegraphics[width = \figwidth]{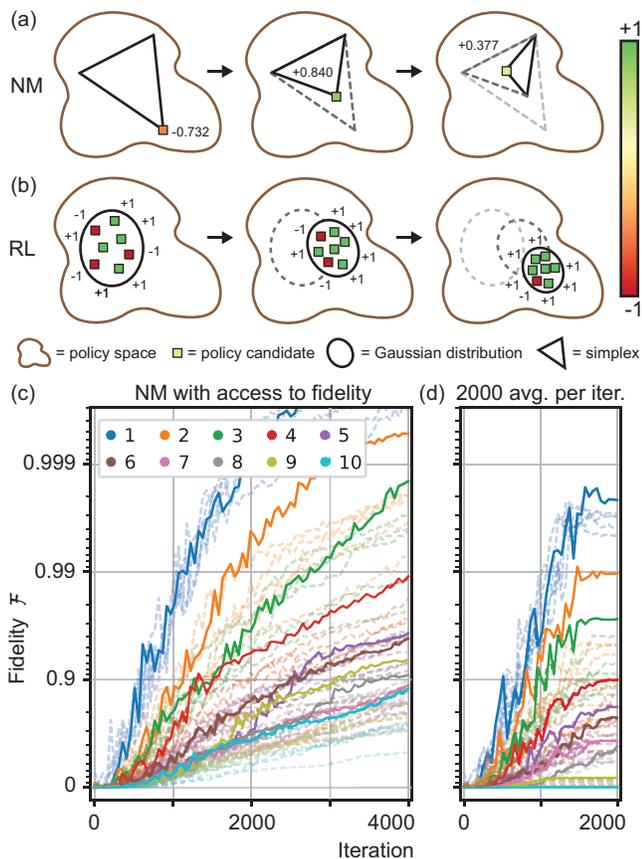}
 \caption{\label{fig7}
Preparation of Fock states $|1\rangle,\, ...,\, |10\rangle$ with Nelder-Mead simplex search.
{\bf (a)} Cartoon depiction of the NM simplex search. One algorithm iteration corresponds to an update of one vertex of a simplex. The color-coded values of the cost function have high resolution, achieved through averaging of many $\pm1$ measurement outcomes. 
{\bf (b)} Cartoon depiction of our RL approach. Every training epoch consists of several episodes executed with different policy candidates that are sampled from a Gaussian distribution. Policy candidates are assigned a low-resolution reward of $\pm1$ based on a single measurement outcome instead of averaging.
{\bf (c)} NM optimization progress with infidelity used as a cost function. The background trajectories correspond to 6 random seeds for each state, solid lines show the trajectory with the highest final fidelity.
{\bf (d)} NM optimization progress with a stochastic cost function obtained by averaging 2000 outcomes of the Fock reward circuit shown in Fig.~\ref{fig2}(a). 
}
\end{figure}

It is instructive to compare the action space exploration strategy of our RL agent to widely used model-free methods. For this comparison, we focus on the Nelder-Mead (NM) simplex search used in many quantum control experiments \cite{Kelly2014, Chen2016, Rol2017}. 

NM and other model-free methods that view quantum control as a standard cost function optimization problem, explore the action space by evaluating the cost function for a set of policy candidates, and using this evaluation to inform the selection of the next candidate. In NM, the latter step is done by choosing a new vertex of the simplex, as illustrated in Fig.~\ref{fig7}(a). The effectiveness of such approach relies on the ability to reliably approximate the cost function landscape by only sampling it at a small subset of points. In general, this is difficult to achieve in high-dimensional action spaces or when the cost function is stochastic. Therefore, such approach requires spending a large part of the sample budget on averaging, which limits the number of policies that it can explore under the constraint of a fixed total sample size of $M_{\rm tot}$ experimental runs. 

On the other hand, in our RL approach every experimental run (episode) is performed with a slightly different policy. These random policy candidates are assigned a stochastic score of $\pm 1$, resulting from the reward measurement outcome. Even though the value of the ``cost function'' is not known to any satisfying accuracy for any of the policy candidates, the acquired information is sufficient to stochastically move the Gaussian distribution of policy candidates towards a more promising region of the action space, as illustrated in Fig.~\ref{fig7}(b). 
In contrast to NM that crucially relies on averaging, our RL agent spends the sample budget to effectively explore a much larger part of the action space.

To confirm this intuition, we quantitatively compare the RL agent to widely used model-free approaches, Nelder-Mead (NM) simplex search and simulated annealing (SA), on the task of Fock state preparation when constrained to the same total sample size of $M_{\rm tot}=4\cdot 10^6$. The results of this comparison are shown  in Fig.~\ref{fig2}(c), revealing that RL indeed significantly outperforms its model-free alternatives in terms of sample efficiency, especially when the effective problem dimension increases, i.e. for higher photon numbers $n$. In the following Sections, we describe the numerical experiments with NM and SA, performed using their  SciPy~1.4.1 implementation \cite{2020SciPy-NMeth}.

\subsection{Nelder-Mead simplex search}

To ensure a fair comparison of NM with RL, we perform hyperparameter tuning for NM, and display the best of 6 independent optimization runs for each problem setting. Given the simplicity of the NM heuristic with its small number of hyperparameters, we believe that the performed tuning is exhaustive and that no further significant improvements are possible.

First, we study the performance of NM when it is given direct access to fidelity on the task of Fock state preparation. We initialize the control circuits with random parameters whose magnitude is swept  to optimize the NM performance, as it is known to be sensitive to the simplex initialization. We find that the optimal initialization is similar to that in RL, and corresponds to random initial circuits that do not significantly deviate the oscillator state from vacuum. With this choice, the convergence of NM is shown in Fig.~\ref{fig7}(c). It exhibits fast degradation with increasing photon number $n$.
Next, we study the performance of NM in the presence of measurement sampling noise. We constrain NM to the same total sample size $M_{\rm tot}=4\cdot 10^6$ as used for RL, and optimally split the sample budget between algorithm iterations and averages per iteration to maximize the final performance. The convergence of NM with 2000 averages per iteration is shown in Fig.~\ref{fig7}(d), and can be directly compared to the RL results in Fig.~\ref{fig2}(b), clearly showing the advantage of RL in stochastic setting.

\subsection{Simulated annealing}

\begin{figure}[t]
 \includegraphics[width = \figwidth]{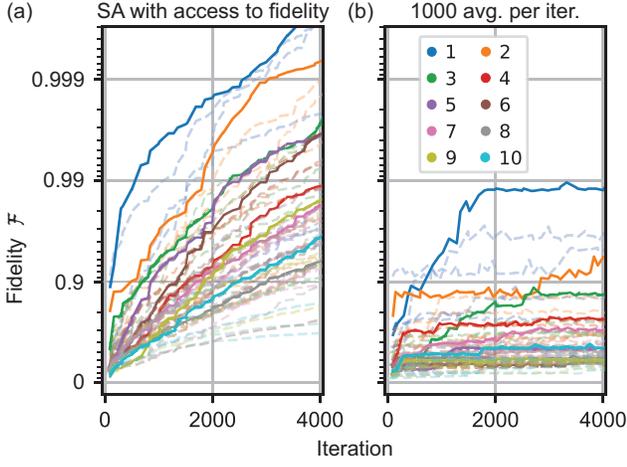}
 \caption{\label{fig8}
Preparation of Fock states $|1\rangle,\, ...,\, |10\rangle$ with simulated annealing.
{\bf (a)} SA optimization progress with infidelity used as a cost function. The background trajectories correspond to 6 random seeds for each state, solid lines show the trajectory with the highest final fidelity.
{\bf (b)} SA optimization progress with a stochastic cost function obtained by averaging 1000 outcomes of the Fock reward circuit shown in Fig.~\ref{fig2}(a). 
}
\end{figure}

We use simulated annealing with Cauchy-Lorentz visiting distribution and without local search on accepted locations, which is a similar version to the recent experiment \cite{Baum2021}. We performed extensive tuning of hyperparameters, including the magnitude of the randomly initialized control circuit parameters, parameters of the visiting distribution, as well as initial and final temperatures. The optimization results with the best choice of hyperparameters are shown in Fig.~\ref{fig8}, where for each optimization trajectory we only display the best fidelity of every 100 consecutive iterations to reduce the plot clutter resulting from the periodic restarts of the annealing. 

With direct access to fidelity, as shown  in Fig.~\ref{fig8}(a), the convergence of SA is similar to NM, and is significantly slower than the RL agent even when the agent does not have access to fidelity. Next, we replace the fidelity with its estimator based on 1000 runs of the Fock reward circuit. This number of runs per cost function evaluation is tuned to achieve the highest performance under the constrained total sample size of $M_{\rm tot}=4\cdot 10^6$. In such stochastic setting, the performance of SA drops significantly, as shown in Fig.~\ref{fig8}(b),  and is worse than that of both NM and RL. 

\section{Variance of the fidelity estimator \label{sec: variance}}
Variance of the estimator \eqref{fidelity_estimator_wigner} is given by
\begin{align}
{\rm Var} =& \underset{\alpha\sim P}{\mathbb{E}}\,\underset{\psi}{\mathbb{E}}\bigg[\bigg(\frac{2}{P(\alpha)}\Pi_{\alpha}\,W_{{\rm target}}(\alpha)\bigg)^{2}\bigg]\nonumber\\
&-\bigg(\underset{\alpha\sim P}{\mathbb{E}}\,\underset{\psi}{\mathbb{E}}\bigg[\frac{2}{P(\alpha)}\Pi_{\alpha}\,W_{{\rm target}}(\alpha)\bigg]\bigg)^{2}\\
 = &\int\frac{4}{P(\alpha)}W_{{\rm target}}^{2}(\alpha)d\alpha-{\cal F}^{2}, \label{variance}
\end{align}
where we made the simplifications $\Pi_{\alpha}^{2}=1$ and $\underset{\alpha\sim P}{\mathbb{E}}\big[...\big]=\int\big[...\big]P(\alpha)d\alpha$.

We now use variational calculus to find $P(\alpha)$ that minimizes \eqref{variance} with the constraint $\int P(\alpha)d\alpha=1$. The variational derivative is given by
\begin{align}
\delta({\rm Var})=\int\bigg[c-\frac{4}{P^{2}(\alpha)}W_{{\rm target}}^{2}(\alpha)\bigg]\delta P(\alpha)d\alpha,
\end{align}
where $c$ is the Lagrange multiplier for the constraint. From this we find that the optimal sampling distribution satisfies
$P(\alpha)\propto|W_{{\rm target}}(\alpha)|$
and the minimal variance is 
\begin{align}
{\rm min}\{{\rm Var}\}  = 4\left(\int |W_{{\rm target}}(\alpha)|d\alpha\right)^2-{\cal F}^{2}.
\end{align}

We considered the sampling problem in which $N_m=1$ parity measurement is done per phase space point, and in such setting we found an optimal sampling distribution independent of the state that is being characterized -- a rather convenient property for the online training, since the actual prepared state is not known (only the target state is known). 
We can consider a different problem, in which both $W(\alpha)$ and $W_{\rm target}(\alpha)$ are known, and where the goal is to compute the fidelity integral  \eqref{integral} through Monte Carlo phase space sampling. 
This can be relevant, for instance, in a simulation, as an alternative to computing the integral through the Riemann sum.
In such setting, the optimal condition for the variance is modified to $P(\alpha)\propto|W(\alpha)W_{{\rm target}}(\alpha)|$. If, in addition, the fidelity is known in advance to be close to 1, i.e. $W(\alpha)\approx W_{{\rm target}}(\alpha)$, then the optimal sampling distribution becomes $P(\alpha)\propto W_{{\rm target}}^{2}(\alpha)$. The latter does not depend on the state that is being characterized, and therefore it can also be used in the online setting, as was proposed in \cite{Flammia2011, DaSilva2011}. However, such sampling distribution is going to be optimal only in the limit $N_m\gg1$. 

In general, consider fidelity estimation based on $N_\alpha$ phase space points and $N_m$ parity measurements per point, such that the total number of measurements $N=N_\alpha N_m$ is fixed. Under this condition, the optimal choice is $N_\alpha=N$, $N_m=1$ (adopted in this work), in which case the distribution $P(\alpha)\propto|W_{{\rm target}}(\alpha)|$ is optimal. However, due to various hardware constrains, e.g. small memory of the FPGA controller, in some experiments it might be preferred to limit $N_\alpha=C$ and compensate for it by accumulating multiple measurements in each phase space point, i.e. $N_m=N/C\gg1$. Under such constraints, the optimal sampling corresponds to $P(\alpha)\propto W_{{\rm target}}^2(\alpha)$.

\section{Other reward measurement schemes \label{sec:other systems}}

In this Appendix, we describe how our approach can be adapted to control of other physical systems, focusing specifically on the design of probabilistic reward measurement schemes.

\subsection{State preparation in trapped ions}

Universal control of a motional state of a trapped ion can be achieved by utilizing the ion's internal electronic levels as ancilla qubit \cite{Leibfried2003, Bruzewicz2019}. Control policies are typically produces with GRAPE, but modular constructions also exist \cite{Kneer1998}. Regardless of the  control circuit parameterization, our RL approach can be used for model-free learning of its parameters. Here, we propose a reward circuit that can be used for such learning in trapped ions, based on the characteristic function.

The symmetric characteristic function of a continuous-variable system is defined as $C(\alpha)=\langle D(\alpha) \rangle$ \cite{Haroche2006}. It is equal to the 2D Fourier transform of the Wigner function, and is therefore tomographically complete and can be used to construct the fidelity estimator similar to Eq.~\eqref{fidelity_estimator_wigner}:
\begin{align}
{\cal F} &=\frac{1}{\pi}\int d^{2}\alpha\,C(\alpha)\,C^*_{{\rm target}}(\alpha)\\
&=\frac{1}{\pi}\underset{\alpha\sim P}{\mathbb{E}}\,\underset{\psi}{\mathbb{E}}\bigg[\frac{1}{P(\alpha)}D({\alpha})\,C^*_{{\rm target}}(\alpha)\bigg],
\label{fidelity_estimator_characteristic_function}
\end{align}
where $P(\alpha)$ is the phase space sampling distribution. In trapped ions the characteristic function can be measured with phase estimation of the unitary displacement operator \cite{Fluhmann2019,Fluhmann2019a}. 

\begin{figure}[h!]
\includegraphics[width = \figwidth]{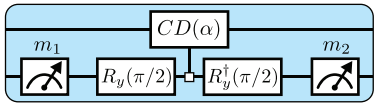}
\caption{Reward circuit for learning preparation of arbitrary symmetric states of a continuous-variable system, based on the characteristic function. \label{fig9}}
\end{figure}

For simplicity, we focus on symmetric states whose characteristic function is real (e.g. Fock states, GKP states), although the procedure can be generalized to asymmetric states. In this case, the reward circuit is similar to Wigner reward, and is shown in Fig.~\ref{fig9}.
The conditional displacement gate $CD(\alpha)$, required for such a reward circuit, is typically called ``internal state dependent force'' in trapped ions community. Note that it was also recently realized in circuit QED \cite{Campagne-Ibarcq2020, Eickbusch2021}.

\subsection{Multi-qubit systems}

Universal control of a system of $n$ qubits with Hilbert space of dimension $d=2^n$ can be achieved with various choices of control circuits that can be tailored to the specific physical layout of the device. We refer to the literature on variational quantum algorithms for more details \cite{Cerezo2020}. Here, we focus instead on the reward measurement schemes. There exists a large body of work on quantum state certification in the multi-qubit systems \cite{Kliesch2021}. Our RL approach greatly benefits from this work, since state certification protocols can be directly converted into probabilistic reward measurement schemes for state preparation control problem. Moreover, some state certification protocols are directly linked to fidelity estimation, which allows to construct reward measurement schemes satisfying the condition $\mathbb{E}[R]=f({\cal F})$, where $f$ is a monotonously increasing function of fidelity.  Here, we propose a stabilizer reward built on the stabilizer state certification protocol \cite{Kliesch2021} and a reward for preparation of arbitrary $n$-qubit states based on the characteristic function. 

\subsubsection{Stabilizer states}

Consider a stabilizer group ${\cal S}=\{I,S_{1},...,S_{d-1}\}$ and a corresponding parameterized set of POVM elements $\{\Omega_{k}\}$ which consists of projectors $\Omega_{k}=\frac{1}{2}(I+S_{k})$ onto the +1 eigenspace of each stabilizer, except for the trivial stabilizer $I$. We sample the parameter $k=1,...,d-1$ uniformly with probabilities $P({k})=\frac{1}{d-1}$, and with the associated identical reward scale $R_{k}=1$. The reward of $\pm1$ is issued based on the stabilizer measurement outcome. A straightforward calculation shows that in this case the expectation of reward satisfies $\mathbb{E}[R]=\frac{2^{n}{\cal F}-1}{2^{n}-1}$, and therefore it also automatically satisfies the condition \eqref{condition}. Note the difference from the GKP state preparation example considered in Section~\ref{sec:Stabilizer states}, where the stabilizer group was infinite and we considered sampling of only the generators of this group, which does not lead to a simple connection between $\mathbb{E}[R]$ and $\cal F$.

\subsubsection{Arbitrary states}
Stabilizer reward is only applicable to a restricted family of states. To construct a reward measurement scheme applicable to arbitrary states, we need to choose a tomographically complete set of POVM elements. The simplest such scheme is based on the Pauli group, where the fidelity estimator can be constructed based on the measurements of $d^2$ possible $n$-fold tensor-products $G_k$ of single-qubit Pauli operators \cite{Flammia2011}. Instead of sampling points $\alpha$ in the continuous phase space, in this case we sample indexes $k$ of the Pauli operators from a discrete set $\{k=1,...,d^2\}$ with probability distribution $P(k)$. Denoting the characteristic function as $C(k)=\langle G_k \rangle$, we obtain an estimator
\begin{align}
{\cal F} &= \frac{1}{d}\sum_k C(k) C_{\rm target}(k)\\
&=\frac{1}{d}\, \underset{k\sim P}{\mathbb{E}}\, \underset{\psi}{\mathbb{E}} \left[ \frac{1}{P(k)} G_k \,C_{\rm target}(k)\right],
\end{align}
Given the estimator above, the reward circuit consists of a measurement of the sampled Pauli operator.

\section{Learning gates for encoded qubits\label{gates}}

The tools demonstrated for quantum state preparation in Section~\ref{results} are applicable for learning more general quantum operations that map an input subspace of the state space to the target output subspace. For example, consider a qubit encoded in oscillator states $\{|\pm Z_L\rangle\}$ which serve as logical $Z$ eigenstates. Learning a gate $U_{\rm target}$ on this logical qubit amounts to finding an operation that simultaneously implements the state transfers $|\pm Z_L\rangle \to U_{\rm target} |\pm Z_L\rangle$ and which extends to logical qubit subspace by linearity. However, the reward circuits introduced in Section~\ref{results} will result in the final state equal to the target up to an arbitrary phase factor, hence it is insufficient to only use the set $\{|\pm Z_L\rangle\}$ during the training. To constrain the phase factor, we extend this set to include all cardinal points $\{|\pm X_L\rangle, |\pm Y_L\rangle, |\pm Z_L\rangle\}$ on the logical Bloch sphere.

The training process for a gate is a straightforward generalization of the training for state preparation depicted in Fig.~\ref{fig1}, as summarized below:

{\bf 1.}  Sample initial state $|\psi_0\rangle \in \{|\pm X_L\rangle, |\pm Y_L\rangle, |\pm Z_L\rangle\}$. Start the episode by preparing this state. 

{\bf 2.} Run the episode by applying $T$ steps of the control circuit, resulting in a state $|\psi_T\rangle$.

{\bf 3.} Apply a reward circuit to state $|\psi_T\rangle$, with the target state given by $|\psi_{\rm target}\rangle = U_{\rm target}|\psi_0\rangle$. 

\begin{figure}[t]
 \includegraphics[width = \figwidth]{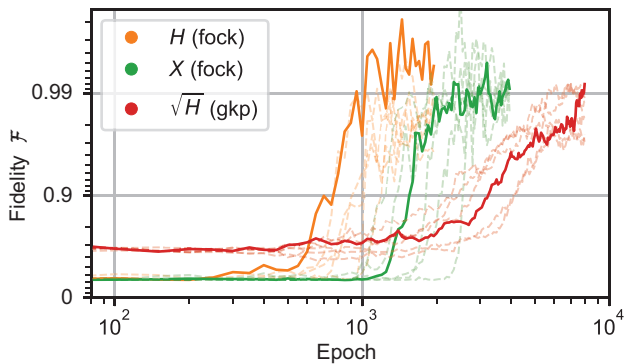}
 \caption{\label{fig10}
Learning gates for logical qubits encoded in an oscillator. The agent is trained to produce a Hadamard $H$ and Pauli $X$ gates on the Fock qubit, and a non-Clifford $\sqrt{H}$ gate on the GKP qubit. Average gate fidelity is used as an evaluation metric. The background trajectories correspond to 6 random seeds for each gate, solid lines show the trajectory with the highest final fidelity.
}
\end{figure}

Here, we demonstrate learning of logical gates for the Fock encoding with $|+Z\rangle = |0\rangle$ and $|-Z\rangle=|1\rangle$, and for the GKP encoding with $\Delta=0.3$. In these numerical experiments, we sample a new initial state every epoch, and use the same state for all batch members within the epoch (preparation of the initial states can be learned beforehand). We use ideal SNAP-displacement control circuit, as shown in Fig.~2(a), and a Wigner reward circuit as shown in Fig.~4(a) with a single phase space point and a single measurement per policy candidate. The choice of training hyperparameters is summarized in the Supplementary Material \cite{SuppMat}.

The training results are displayed in Figure~\ref{fig10}, for the Hadamard $H$ and Pauli $X$ gates on the Fock qubit, and a non-Clifford $\sqrt{H}$ gate on the GKP qubit. We use average gate fidelity \cite{Nielsen2002} as an evaluation metric. These results show that stable convergence is achieved in such QOMDP despite an additional source of randomness due to the sampling of initial states. The total number of experimental realizations used by the agent is $10^6$, $2\cdot 10^6$ and $4\cdot 10^6$ for the $H$, $X$ and $\sqrt{H}$ gates respectively.

In the future work, error amplification technique based on gate repetitions, such as randomized benchmarking (RB), can be incorporated to increase the SNR of the reward, similarly to how it is done in other quantum control demonstrations \cite{Kelly2014, Werninghaus2021}. However, this technique could be modified in the spirit of our approach, to use a single experimental realization of a random RB sequence as one episode, instead of averaging them to suppress the stochasticity of the cost function.

\bibliographystyle{apsrev_longbib}
\bibliography{lib}

\newpage

\suppmat

\title{Supplementary Material \\[1ex] \large
``Model-Free Quantum Control with Reinforcement Learning''}

\author{V. V. Sivak}
\affiliation{Department of Applied Physics, Yale University, New Haven, CT 06520, USA}

\author{A. Eickbusch}
\affiliation{Department of Applied Physics, Yale University, New Haven, CT 06520, USA}

\author{H. Liu}
\affiliation{Department of Applied Physics, Yale University, New Haven, CT 06520, USA}

\author{B. Royer}
\affiliation{Department of Physics, Yale University, New Haven, CT 06520, USA}

\author{I. Tsioutsios}
\affiliation{Department of Applied Physics, Yale University, New Haven, CT 06520, USA}

\author{M. H. Devoret}
\affiliation{Department of Applied Physics, Yale University, New Haven, CT 06520, USA}

\maketitle

\onecolumngrid

\renewcommand{\thesection}{S\arabic{section}}   
\renewcommand{\thetable}{S\arabic{table}}   
\renewcommand{\thefigure}{S\arabic{figure}}
\renewcommand{\theequation}{S\arabic{equation}}

\setcounter{equation}{0}
\setcounter{figure}{0}

\section{Introduction to Proximal Policy Optimization}

In this section, we present a brief derivation of the Proximal Policy Optimization algorithm (PPO) \cite{Schulman2017SM}. For simplicity, we consider a fully-observable MDP, but the results can be generalized to partially-observable MDPs.

The performance measure $J$ in reinforcement learning is the expected return $R(\tau)$ of the trajectory (episode) $\tau$ generated while following the policy $\pi_{\theta}$:
\begin{equation}
J(\pi_{\theta})=\underset{\tau\sim\pi_{\theta}}{\mathbb{E}}[R(\tau)]=\sum_{\tau}R(\tau)P(\tau|\theta), \label{eq:performance measure}
\end{equation}
where the probability $P(\tau|\theta)$ of the trajectory $\tau$ under the policy $\pi_{\theta}$ is given by the product over all time steps of the conditional probabilities $\pi_{\theta}(a_{t}|s_{t})$ of choosing the action $a_t$ in the state $s_t$, and ${\cal T}(s_{t+1}|s_{t},a_{t})$ of environment transition $s_t\to s_{t+1}$ given this choice of action:
\begin{equation}
P(\tau|\theta)=\prod_t {\cal T}(s_{t+1}|s_{t},a_{t})\pi_{\theta}(a_{t}|s_{t}).
\end{equation}
With this expression substituted in Eq.~\eqref{eq:performance measure}, we can compute the gradient of the performance measure
\begin{align}
\nabla_{\theta}J(\pi_{\theta}) & =\sum_{\tau}R(\tau)\nabla_{\theta}P(\tau|\theta)=\sum_{\tau}R(\tau)P(\tau|\theta)\nabla_{\theta}\log P(\tau|\theta)\\
 & =\underset{\tau\sim\pi_{\theta}}{\mathbb{E}}\big[R(\tau)\nabla_{\theta}\log P(\tau|\theta)\big]=\sum_{t}\underset{\tau\sim\pi_{\theta}}{\mathbb{E}}\bigg[R(\tau)\nabla_{\theta}\log\pi_{\theta}(a_{t}|s_{t})\bigg]. \label{gradient}
\end{align}

Note that the environment transition function ${\cal T}(s_{t+1}|s_{t},a_{t})$ has dropped out because it is independent of the policy parameters -- a crucial feature enabling model-free learning. The performance gradient \eqref{gradient} has a natural form of the sum of policy gradients over all time-steps of the trajectory, weighted by the return of the trajectory $R(\tau)$. Such gradient will increase the probabilities of actions that caused high return in the past experiences, and decrease the probabilities of actions that caused low return. However, as explained in detail in \cite{Schulman2015SM}, such weighting with $R(\tau)$ is sub-optimal in terms of the estimator variance. For instance, it propagates the influence of the rewards received prior to applying a given action $a_t$ on the score of this action, which indeed seems counter-intuitive. A better weighting can be obtained by replacing the full trajectory return $R(\tau)$ with the partial return $R(\tau; s_t, a_t)$ accumulated in trajectory $\tau$ after visiting the state $s_t$ and taking the action $a_t$. Such replacement preserves the unbiased nature of the estimator, but allows to reduce its variance. Further improvement can be obtained by subtracting the state-dependent baseline $b(s_t)$, which helps to ensure that good (relative to the baseline) actions have positive weight, while bad actions have negative weight. The baseline can be any function that only depends on the state $s_t$, but the optimal baseline would satisfy the condition $b(s_t)={\mathbb{E}_{\tau\sim\pi_{\theta}}}[R(\tau;s_t)]$, where $R(\tau;s_t)$ is the partial return accumulated in trajectory $\tau$ after visiting the state $s_t$ and averaged over all possible actions in that state. In practice, since the optimal baseline is not known in advance, it is represented with a {\it value} neural network $V_{\theta'}(s_{t})$ whose parameters $\theta'$ are learned concurrently with parameters $\theta$ of the policy network. Incorporating these improvements leads to the following weighing factor for the policy gradients in Eq.~\eqref{gradient}, known as the empirical advantage function
\begin{align}
A(\tau; s_t, a_t)=R(\tau; s_t, a_t)-V_{\theta'}(s_{t}). \label{advantage}
\end{align}

When using the advantage estimator \eqref{advantage} in place of the empirical return $R(\tau)$ in \eqref{gradient}, the performance gradient becomes
\begin{align}
\nabla_{\theta}J(\pi_{\theta})=\nabla_{\theta}\sum_{t}L_{t}^{PG},\qquad L_{t}^{PG}=\hat{\mathbb{E}}\left[A(\tau;s_t, a_t)\log\pi_{\theta}(a_{t}|s_{t})\right],
\end{align}
where $L_{t}^{PG}$ is the per-time-step policy-gradient loss that can be used with automatic differentiation, and $\hat{\mathbb{E}}[...]$ is an empirical average over a finite batch of $B$ trajectories. 

In the actor-critic methods \cite{Sutton2017SM}, which PPO also belongs to, the value function $V_{\theta'}$ (critic) is learned concurrently with the policy $\pi_{\theta}$ (actor) to predict the partial return $R(\tau; s_t)$. Typically, this is achieved with a simple quadratic loss $L_{t}^{V}=[R(\tau;s_t)-V_{\theta'}(s_{t})]^{2}$. Policy-gradient loss $L_{t}^{PG}$ and value-function loss $L_{t}^{V}$ are combined to compute the total gradient which is passed on to the optimizer.

What was described so far is the basic working principle of the REINFORCE algorithm \cite{Sutton1999SM}. A convenient feature of the simple policy gradient is that we can use the first order optimizers such as stochastic gradient descent (SGD) or Adam \cite{Kingma2014SM} to minimize the loss function. However, it was found that such policy optimization in a high dimensional space of neural network parameters is often unstable -- there can be drastic jumps in the policy performance even with the small changes of the parameters $\theta$ of the policy. The trust-region policy optimization algorithm (TRPO) \cite{Schulman2015aSM}, which is a precursor to PPO, attempted to cure this issue by using an expensive second order optimization within the trust region where the Kullback–Leibler divergence between the old and the new policy is constrained. PPO emerged as an attempt to get the best of both worlds: efficiency of the first order optimizers and guarantee that the policy will not make any catastrophic jumps. It achieves this by constructing a special (although very simple) loss function that does not incentivize the optimizer to deviate the new policy far from the old one. 

To derive the PPO loss function, we first rewrite the per-time-step gradient of $L_t^{PG}$ using importance sampling
\begin{equation}
\underset{\tau\sim\pi_{\theta}}{\hat{\mathbb{E}}}\big[\frac{\nabla_{\theta}\pi_{\theta}(a_{t}|s_{t})}{\pi_{\theta}(a_{t}|s_{t})}A(\tau; s_t,a_t)\big]=\underset{\tau\sim\pi_{{\rm old}}}{\hat{\mathbb{E}}}\big[\frac{\pi_{\theta}(a_{t}|s_{t})}{\pi_{\theta_{{\rm old}}}(a_{t}|s_{t})}\frac{\nabla_{\theta}\pi_{\theta}(a_{t}|s_{t})}{\pi_{\theta}(a_{t}|s_{t})}A(\tau;s_t,a_t)\big]=\underset{\tau\sim\pi_{{\rm old}}}{\hat{\mathbb{E}}}\big[\frac{\nabla_{\theta}\pi_{\theta}(a_{t}|s_{t})}{\pi_{\theta_{{\rm old}}}(a_{t}|s_{t})}A(\tau;s_t,a_t)\big],
\end{equation}
which leads to the per-time-step loss contribution $L_{t}=\hat{\mathbb{E}}\big[A(\tau;s_t,a_t)\frac{\pi_{\theta}(a_{t}|s_{t})}{\pi_{\theta_{{\rm old}}}(a_{t}|s_{t})}\big]$.

If a small change in the policy parameters $\theta$ causes policy to differ significantly from the old policy with parameters $\theta_{\rm old}$, the importance ratio
$\frac{\pi_{\theta}(a_{t}|s_{t})}{\pi_{\theta_{{\rm old}}}(a_{t}|s_{t})}$ will deviate significantly from 1. PPO simply clips the importance ratio to the range $(1-\epsilon,1+\epsilon)$ (where typically $\epsilon\sim0.2$), leading to the new per-time-step loss
\begin{equation}
L_{t}^{PPO}=\hat{\mathbb{E}}\left[{\rm min}\bigg(\frac{\pi_{\theta}(a_{t}|s_{t})}{\pi_{\theta_{{\rm old}}}(a_{t}|s_{t})}A(\tau;s_t,a_t),\,{\rm clip}\bigg[\frac{\pi_{\theta}(a_{t}|s_{t})}{\pi_{\theta_{{\rm old}}}(a_{t}|s_{t})}\bigg]A(\tau;s_t,a_t)\bigg)\right],
\end{equation}
replacing $L_t^{PG}$ loss. With such modification, if a certain policy update $\theta_{{\rm old}}\to\theta$ attempts to reduce the loss by making the importance ratio deviate significanty from 1, it will not be able to achieve this because the importance ratio will be clipped and the loss will not benefit from such an update. Therefore, importance ratio clipping removes the incentive for such changes, although it does not strictly guarantee that they will not happen. Empirically, this leads to significantly improved stability of the training, which is especially relevant in stochastic environments.

\section{Implementation of the training environment}
We realize a custom training environment following TensorFlow Agents interface \cite{Hafner2017SM}. TF-Agents is an open-source library for reinforcement learning which provides reliable implementations of several popular algorithms including PPO. The TensorFlow implementation of both the custom training environment and the agent was efficiently accelerated with the graphics processing unit NVIDIA Tesla V100. Given the computational demand of our task (tuning hyperparameters, exploring different fidelity estimators, training for different quantum states, collecting statistics over multiple random seeds,  etc), the GPU acceleration of the simulation can be acknowledged as the most significant factor in the success of this numerical project. 
As an example, in the training for Fock state preparation described in Section IV A, the wall clock time of the quantum simulation alone (excluding the neural network update time) is 13 minutes in total for 4000 epochs consisting of 1000 episodes each.

Several most significant factors contributing to the numerical complexity of the project:

\begin{enumerate}
\item The Hilbert space of the oscillator is truncated at $N=100$ states in the photon number basis, and in product with the ancilla qubit this leads to $200$-dimensional vectors representing the quantum states (for the GKP states, the truncation is increased to $N=200$). The operators on the joint Hilbert space are $200\times200$ complex-valued matrices represented in the single-precision floating point format to speed up the computation.

\item  At each time step $t=1,...,T$ of the trajectory, the agent predicts a {\it new} parameterization of the control circuit, and thus the operators cannot be pre-computed and stored in memory and instead need to be computed on-the-fly. For the displacement operators $D(\alpha)$, instead of performing expensive matrix exponentiation at every time-step, we implement this subroutine efficiently using the Baker–Campbell–Hausdorff (BCH) formula for matrix exponential, see Section~\ref{BCH}. For the finite-duration $\rm SNAP_\tau(\varphi)$ gate, we use a closed-form approximate model vectorizable on the GPU, instead of time-domain integration of the Schrödinger equation, see Section~\ref{SNAP}.

\item Model-free reinforcement learning, and especially policy-gradient algorithms, is known to have poor sample efficiency, which stems from the need to collect new training dataset after each update of the policy. In our state preparation examples, each training requires tens of millions of episodes. We implemented an efficient vectorized quantum trajectory simulator on the GPU, which allows to collect batches of $B\sim 1000$ episodes in parallel. In addition to unitary dynamics, it allows to simulate asynchronous quantum jumps (not used in this work).
\end{enumerate}

\subsection{Efficient implementation of the displacement operator \label{BCH}}
In addition to utilizing GPU acceleration and vectorization for the batch, we can further take advantage of the structure of the displacement operator $D(\alpha)=\exp(\alpha a^\dagger -\alpha^* a)$ to customize the matrix exponentiation routine. We first rewrite $D(\alpha)$ using the position $x=(a+a^\dagger)/\sqrt{2}$ and momentum $p=i(a^\dagger-a)/\sqrt{2}$ Hermitian operators:
\begin{align}
D(\alpha) = e^{-i\sqrt{2}{\rm Re}(\alpha)p + i\sqrt{2}{\rm Im}(\alpha) x}.
\end{align}
This expression can be further transformed using the BCH formula $e^{A+B}=e^{A}e^{B}e^{-\frac{1}{2}[A,B]}$, which is exact in this case because both $x$ and $p$ commute with their commutator $[x,p]=i$: 
\begin{align}
D(\alpha)=e^{i\sqrt{2}{\rm Im}(\alpha)x}e^{-i\sqrt{2}{\rm Re}(\alpha)p}e^{-i{\rm Im}(\alpha){\rm Re}(\alpha)}.
\end{align} 
 
\begin{figure}[h!]
 \includegraphics[width = \figwidth]{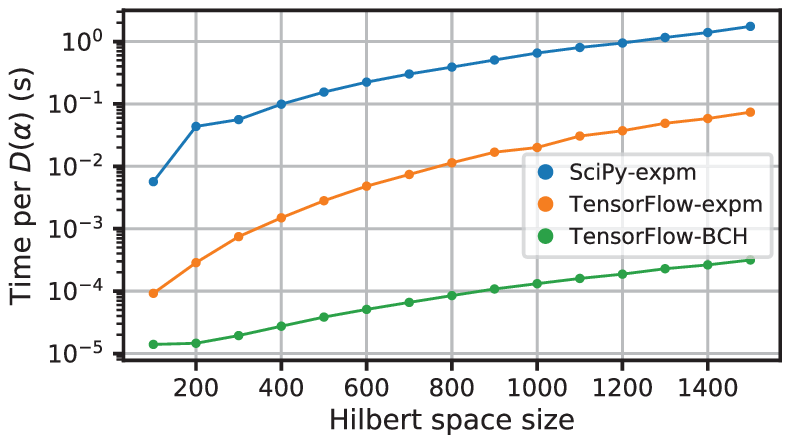}
 \caption{\label{fig_S1} Benchmarking of different implementations of the displacement operator $D(\alpha)$. A batch of $B=100$ displacement operators with random amplitudes is generated using matrix exponential (expm) from SciPy library, and compared to TensorFlow implementations using both the matrix exponential (expm) and BCH approach described in Section~\ref{BCH}. All methods contain additional speedup due to vectorization of the batch. The host computer runs on Intel Xeon Gold 6240 processor and has NVIDIA Tesla V100 GPU.}
\end{figure}

Given this form, $D(\alpha)$ can be computed efficiently by pre-diagonalizing $x$ and $p$ operators in the beginning of the training as
\begin{align}
x=U_{x}\,{\rm diag}(x)\,U_{x}^{\dagger}\qquad p=U_{p}\,{\rm diag}(p)\,U_{p}^{\dagger},
\end{align}
and then performing only element-wise diagonal matrix exponentiation at each time step 
\begin{align}
D(\alpha)=U_{x}\,e^{i\sqrt{2}{\rm Im}(\alpha){\rm diag}(x)}\,U_{x}^{\dagger}\,U_{p}\,e^{-i\sqrt{2}{\rm Re}(\alpha){\rm diag}(p)}\,U_{p}^{\dagger}\,e^{-i{\rm Im}(\alpha){\rm Re}(\alpha)}.
\end{align}
The complexity of this algorithm is $O(N^{3})$ due to matrix multiplication, which is similar to the complexity of the general matrix exponentiation, but the pre-factor is $10-100$ times smaller, see Fig.~\ref{fig_S1} for comparison.

\subsection{Simulation of the finite-duration SNAP \label{SNAP}}

To simulate the effect of the SNAP gate in the real experiment, we need to build an approximate model of the partially-selective qubit pulse, but it is important to realize that the agent is not aware of it by construction. We require that this model captures the essence of the finite-duration pulse and that it can be efficiently implemented numerically, but it does not necessarily have to be very accurate. In general, the problem of driving the qubit with arbitrary time-dependent field is not analytically solvable. The formal solution involves the time-ordered exponential, which can be numerically evaluated with expensive time-domain integration of the Schrödinger equation assuming a certain pulse shape, but this is not an efficient solution for our purpose. Instead, we would like to obtain a closed-form model for the unitary ${\rm SNAP}_{\tau}(\varphi)$ which can be vectorized on the GPU. 

The perfect SNAP gate is equivalent to

\begin{equation}
{\rm SNAP}(\varphi)=\sum_{n}|n\rangle\langle n|\otimes R_{\pi-\varphi_{n}}(\pi)R_{0}(\pi)\label{eq:snap with ideal selective pulse}
\end{equation}
where $R_{\phi}(\vartheta)=\exp\big(-i\frac{\vartheta}{2}(\cos\phi\,\sigma_{x}+\sin\phi\,\sigma_{y})\big)$
is the qubit rotation operator. Such decomposition is inspired by
the availability of well-controlled selective qubit rotations in the
strong dispersive limit of circuit QED.

In general, given that the dispersive coupling $H_{c}/h=\frac12\chi a^{\dagger}a \sigma_z$
preserves the oscillator photon number, driving the qubit with arbitrary
near-resonant time-dependent pulse of duration $\tau$ leads to the
joint unitary gate $U=\sum_{n}|n\rangle\langle n|\otimes R_{\phi_{n}}(\vartheta_{n})$
where the gate parameters $\{(\phi_{n},\vartheta_{n})\}_{n=0}^{\infty}$
depend on the pulse composition which in turn depends on the action produced by the agent. 
In experiment, the controlled mapping
from parameters of the pulse waveform to parameters of the unitary
can be achieved in two limiting cases of long $\chi\tau\gg1$ selective
pulses or short $\chi\tau\ll1$ un-selective pulses. The intermediate
case $\chi\tau\sim1$ is hard to treat analytically because of the
absence of small parameters, and hard to simulate with
high accuracy because of the sensitivity to experimental distortions
(unlike in the limiting cases where distortions could be calibrated
out). 

To practically implement the SNAP truncated at $\Phi$ levels, we use
$\Phi$ carrier frequency components $f_{k}=f_{q}-k\chi$ with $k=0,...,\Phi-1$
in the composite qubit pulse, where $f_{q}$ is the qubit frequency.
For simplicity, we assume that pulse components have rectangular envelopes with amplitudes $\Omega_{k}$ and phases $\delta_{k}$. 
Such composite pulse is described by the drive Hamiltonian $H_{d}/h=\frac{1}{2}({\rm Re}[\Omega(t)] \sigma_x + {\rm Im}[\Omega(t)] \sigma_y)$, where $\Omega(t)=\sum_{k=0}^{\Phi-1} \Omega_k e^{2\pi\chi k t i+\delta_k i}$.
After performing the unitary transformation on the total Hamiltonian $H=H_c+H_d$ to eliminate the dispersive term $H_c$,
we obtain the following time-dependent Hamiltonian describing the system evolution during the pulse
\begin{align}
H(t)=\sum_{n=0}^\infty|n\rangle\langle n|\otimes\sum_{k=0}^{\Phi-1}\frac{\Omega_{k}}{2}\bigg[\cos(\Delta_{kn}t+\delta_{k})\sigma_{x}+\sin(\Delta_{kn}t+\delta_{k})\sigma_{y}\bigg],
\end{align}
where $\Delta_{kn}=2\pi\chi(k-n)$. In this sum, the terms with $k=n$
are resonant and thus non-rotating, while all other terms correspond
to detuned driving of transition $n$ with the pulse component $k$
and are thus rotating. 

To obtain a simple closed-form model of the unitary gate $U={\cal T}\exp(-i\int_0^\tau H(t)dt)$ implemented by such pulse, we use the first-order rotating wave approximation (RWA) and replace the time-dependent Hamiltonian $H(t)$ with a constant time-averaged Hamiltonian $\overline{H}=\frac{1}{\tau}\int_0^\tau H(t)dt$. Effectively, this removes the time-ordering operation in the unitary $U=\exp(-i\int_0^\tau H(t)dt)$, leading to
\begin{equation}
U=\sum_{n=0}^{\infty}|n\rangle\langle n|\otimes\exp\bigg\{-i\sum_{k=0}^{\Phi-1}\frac{\Omega_{k}\tau}{2}\bigg[\frac{\sin(\Delta_{kn}\tau+\delta_{k})-\sin(\delta_{k})}{\Delta_{kn}\tau}\sigma_{x}-\frac{\cos(\Delta_{kn}\tau+\delta_{k})-\cos(\delta_{k})}{\Delta_{kn}\tau}\sigma_{y}\bigg]\bigg\}. \label{eq:pulse model}
\end{equation}
This is not a parametric approximation in some small parameter, but it captures the essential effect of the pulse. In particular, it will lead to the leftover entanglement between the qubit and the oscillator after the SNAP gate. 

In the limit $\chi\tau\gg1$ the unitary \eqref{eq:pulse model} simplifies to the selective qubit rotation where each number-split transition is only affected by the resonant pulse component
\begin{equation}
U_{\chi\tau\gg1}=\sum_{n=0}^{\infty}|n\rangle\langle n|\otimes\exp\bigg(-i\frac{\vartheta_{n}}{2}(\cos\phi_{n}\sigma_{x}+\sin\phi_{n}\sigma_{y})\bigg),\label{eq:selective}
\end{equation}
where $\vartheta_{k}=\Omega_{k}\tau/2$, $\phi_{k}=\delta_{k}$ for $k=0,...,\Phi-1$ and $\vartheta_{k}=\phi_{k}=0$ for $k\ge\Phi$
is a simple mapping from the parameters of the pulse $\{(\delta_{k},\Omega_{k})\}_{k=0}^{\Phi-1}$
to the parameters of the unitary $\{(\phi_{k},\vartheta_{k})\}_{k=0}^{\infty}$.

In the short-time limit $\chi\tau\ll1$ the unitary \eqref{eq:pulse model}
yields the un-selective qubit rotation
\begin{align}
U_{\chi\tau\ll1}=I\otimes\exp\bigg(-i\sum_{k=0}^{\Phi-1}\vartheta_{k}[\cos\phi_{k}\sigma_{x}+\sin\phi_{k}\sigma_{y}]\bigg).
\end{align}

We use the unitary \eqref{eq:pulse model} to interpolate between
these two limits and to build a partially-selective ${\rm SNAP}_{\tau}(\varphi)$
gate in the following way. In this gate, the first qubit pulse $R_{0}(\pi)$
can always be done in the fast un-selective manner, and thus we simulate
it as a perfect rotation. The second pulse depends on the action parameters
$\{\varphi_{k}\}_{k=0}^{\Phi-1}$ produced by the agent. We map the action
component $\varphi_{k}$ to the corresponding pulse component assuming
that the pulse is perfectly selective as in \eqref{eq:selective},
$\delta_{k}=\pi-\varphi_{k}$ and $\Omega_{k}=2\pi/\tau$, but end
up applying the partially selective unitary \eqref{eq:pulse model}
with these parameters. Such approximate model allows us to roughly capture
the expected degradation of performance that the protocol would exhibit
in the real experiment, in contrast to other ad-hoc simulations of
control imperfections, such as, for instance, injecting random static
offsets in the qubit rotation matrix.

\section{Simulation parameters and training hyperparameters}

The simulation parameters and training hyperparameters used for state preparation examples in the main text are summarized in Table~\ref{hyperparameters}, and those used for gates on encoded qubits in Appendix E are summarized in Table~\ref{gates hyperparameters}.

\begin{table}
\begin{centering}
\renewcommand{\arraystretch}{1.3}
\begin{tabular}{|c|c|c|c|c|c|}
\hline 
Target state & fock1--10 & cat2 & bin1 & gkp & 
\begin{minipage}[t]{0.10\columnwidth}
fock3\\
(adaptive)
\end{minipage}
\tabularnewline
\hline 
Epochs & $4\cdot10^3$ & 
\begin{minipage}[t]{0.10\columnwidth}
$2\cdot10^4$,\\ $4\cdot10^3$,\\ $10^3$
\end{minipage} & 
\begin{minipage}[t]{0.10\columnwidth}
$2\cdot10^4$,\\ $1\cdot10^4$,\\ $4\cdot10^3$
\end{minipage} & $10^4$ & $2.5\cdot10^4$
\tabularnewline
\hline 
Episodes per epoch & $10^3$ & $10^3$ & 500 & $10^3$ & $10^3$
\tabularnewline
\hline 
Learning rate schedule & 
\begin{minipage}[t]{0.16\columnwidth}
$10^{-3}$, epoch $<500$\\
$10^{-4}$, epoch $\ge500$
\end{minipage} & $10^{-3}$ & $10^{-3}$ & $10^{-3}$ & 
\begin{minipage}[t]{0.17\columnwidth}
$10^{-3}$, epoch $<1000$\\
$10^{-4}$, epoch $\ge1000$
\end{minipage}
\tabularnewline
\hline 
Gradient norm clipping & 1 & 1 & 1 & 1 & 1\tabularnewline
\hline 
Importance ratio clipping & $1\pm0.1$ & $1\pm0.1$ & $1\pm0.2$ & $1\pm0.25$ & $1\pm0.1$\tabularnewline
\hline 
\begin{minipage}[t]{0.20\columnwidth}
Policy \& value networks hidden layers
\end{minipage}
 & \begin{minipage}[t]{0.10\columnwidth}
LSTM(16)\\
Dense(100)\\
Dense(50)
\end{minipage}
 & \begin{minipage}[t]{0.10\columnwidth}
LSTM(12)
\end{minipage} & 
\begin{minipage}[t]{0.10\columnwidth}
LSTM(12)\\
Dense(50)
\end{minipage}
 & \begin{minipage}[t]{0.10\columnwidth}
LSTM(12)
\end{minipage} & \begin{minipage}[t]{0.10\columnwidth}
LSTM(16)\\
Dense(100)\\
Dense(50)
\end{minipage}
\tabularnewline
\hline 
Value prediction loss weight & $5\cdot10^{-3}$ & $5\cdot10^{-3}$ & $5\cdot10^{-3}$ & $5\cdot10^{-3}$ & $5\cdot10^{-3}$\tabularnewline
\hline 
Joint Hilbert space size, $2N$ & 200 & 200 & 200 & 400 & 200
\tabularnewline
\hline 
SNAP truncation, $\Phi$ & 15 & 10 & 15 & 30 & 7
\tabularnewline
\hline 
Time steps, $T$ & 5 & 5 & 8 & 9 & 5
\tabularnewline
\hline 
Reward function & Fock & 
\begin{minipage}[t]{0.10\columnwidth}
Wigner \\
(1, 10, 100\\
pts avg)
\end{minipage} & 
\begin{minipage}[t]{0.10\columnwidth}
Wigner \\
(1, 10, 100\\
pts avg)
\end{minipage} & Stabilizers & Fock
\tabularnewline
\hline 
\end{tabular}
\par
\end{centering}
\caption{Simulation parameters and training hyperparameters used for state preparation examples in the main text.\label{hyperparameters}}
\end{table}

\begin{table}
\begin{centering}
\renewcommand{\arraystretch}{1.3}
\begin{tabular}{|c|c|c|c|}
\hline 
Target gate & $H$ & $X$ & $\sqrt{H}$
\tabularnewline
\hline 
Logical encoding & Fock & Fock & GKP, $\Delta=0.3$
\tabularnewline
\hline 
Epochs & $2\cdot10^3$ & $4\cdot10^3$ & $8\cdot10^3$
\tabularnewline
\hline 
Episodes per epoch & 500 & 500 & 500
\tabularnewline
\hline 
Learning rate schedule & $10^{-3}$ & $10^{-3}$ & $10^{-3}$
\tabularnewline
\hline 
Gradient norm clipping & 1 & 1 & 1
\tabularnewline
\hline 
Importance ratio clipping & $1\pm0.1$ & $1\pm0.1$ & $1\pm0.1$
\tabularnewline
\hline 
\begin{minipage}[t]{0.20\columnwidth}
Policy \& value networks hidden layers
\end{minipage}
 & \begin{minipage}[t]{0.10\columnwidth}
LSTM(12)\\
Dense(50)
\end{minipage}
 & \begin{minipage}[t]{0.10\columnwidth}
LSTM(12)\\
Dense(50)
\end{minipage} & 
\begin{minipage}[t]{0.10\columnwidth}
LSTM(12)\\
Dense(50)
\end{minipage}
\tabularnewline
\hline 
Value prediction loss weight & $5\cdot10^{-3}$ & $5\cdot10^{-3}$ & $5\cdot10^{-3}$
\tabularnewline
\hline 
Joint Hilbert space size, $2N$ & 200 & 200 & 300
\tabularnewline
\hline 
SNAP truncation, $\Phi$ & 15 & 15 & 80
\tabularnewline
\hline 
Time steps, $T$ & 4 & 4 & 1
\tabularnewline
\hline 
Reward function & Wigner (1pt) & Wigner (1pt) & Wigner (1pt)
\tabularnewline
\hline 
\end{tabular}
\par
\end{centering}
\caption{
Simulation parameters and training hyperparameters used for logical gate examples in Appendix E. \label{gates hyperparameters}}
\end{table}

\bibliographystyleSM{apsrev_longbib}
\bibliographySM{SM}

\end{document}